\documentclass[12pt]{article}
\usepackage{amssymb}
\usepackage{graphicx}
\usepackage{sint,cite}

\newcommand{\tr}{\mathrm{tr}}

\newcommand{\beq}{\begin{eqnarray}}
\newcommand{\eeq}{\end{eqnarray}}

\newcommand{\VEV}[1]{\left\langle #1 \right\rangle}

\newcommand{\eq}[1]{Eq.~(\ref{#1})}
\newcommand{\fig}[1]{Fig.~\ref{#1}}
\newcommand{\tab}[1]{Table~\ref{#1}}
\newcommand{\sect}[1]{Section~\ref{#1}}
\newcommand{\plaq}{{\rm Plaq}}
\newcommand{\pq}[1]{\plaq_{#1}}
\newcommand{\poly}{{\rm Poly}}

\def\aqq{\alpha_{\rm qq}}
\def\CF{C_{\rm F}}
\def\CA{C_{\rm A}}
\def\gqq{\bar{g}_\mathrm{qq}}
\def\bqq{b^{\rm(qq)}}
\def\betaqq{\beta_\mathrm{qq}}
\def\Lmin{L^{\rm min}}
\def\Ls{L_{\rm s}}
\def\Lsmin{\Lmin_{\rm s}}
\def\L5min{\Lmin_5}
\newcommand{\onecol}[2]{
        \begin{minipage}[t]{#1}{#2\vfill} \end{minipage}
        }
\def\taui{\tau_{\rm int}}

\begin{document}

\thispagestyle{empty}
\title{{\normalsize
\mbox{} \hfill
\onecol{4.0cm}{\vspace{-1.9cm} WUB/11-16}} \\
\vspace{1cm}
On the phase structure of five-dimensional SU(2) gauge theories with
anisotropic couplings
}

\author{
\normalsize Francesco Knechtli $^{a}$, Magdalena Luz $^{a}$ and Antonio Rago
$^{a,b}$ \\[0.5cm] \normalsize
$^{a}$ Department of Physics, Bergische Universit\"at Wuppertal\\\normalsize
Gaussstr. 20, D-42119 Wuppertal, Germany\\[0.25cm]\normalsize
$^{b}$ School of Computing and Mathematics, University of
Plymouth\\\normalsize Plymouth PL4 8AA, UK\\[0.5cm]
}
\date{}

\maketitle

\begin{abstract}
The phase diagram of five-dimensional SU(2) gauge theories is explored using
Monte Carlo simulations of the theory discretized on a Euclidean lattice
using the Wilson plaquette action and periodic boundary conditions. We
simulate anisotropic gauge couplings which correspond to different lattice
spacings $a_4$ in the four dimensions and $a_5$ along the extra dimension.
In particular we study the case where $a_5>a_4$. We identify a line of first
order phase transitions which separate the confined from the deconfined phase.
We perform simulations in large volume at the bulk phase transition staying in
the confined vacuum. The static potential measured in the hyperplanes
orthogonal to the extra dimension hint at dimensional reduction. We also
locate and analyze second order phase transitions related to breaking of the
center along one direction.
\end{abstract}


\newpage

\section{Introduction}
\label{sect:introduction}

Our interest in studying five-dimensional gauge theories comes from extensions
of the Standard Model called Gauge-Higgs unification. The idea is to identify
the Higgs boson with (some of) the extra dimensional components of the gauge
field. Since five-dimensional gauge theories are non-renormalizable\footnote{
Equivalently one can say that five-dimensional gauge theories are trivial.}
an ultra-violet cut-off is mandatory. The lattice regularization provides such
a gauge-invariant cut-off (the inverse lattice spacing) and is therefore a
natural setup to study these theories. 

Here we consider the five-dimensional pure SU(2) gauge theory and investigate
its non-perturbative phase diagram through Monte Carlo simulations. The theory
is discretized on a Euclidean space-time lattice. The seminal work done
by Creutz \cite{Creutz:1979dw} using the Wilson plaquette gauge action 
\cite{Wilson:1974sk} demonstrated the existence of (at least)
two bulk phases in the non-perturbative phase diagram of the theory in infinite
volume. There is a confined phase at large values of the gauge coupling and a
deconfined phase when the gauge coupling is small. 

The question we are after
is where dimensional reduction from five to four dimensions occurs. Possible
mechanisms, which have been proposed are compactification or
localization. In compactification models the size of the extra dimension is
small and the system is described by an effective four-dimensional low energy
theory valid for energies much
below the inverse compactification radius. The existence of a minimal physical
size of the extra dimension, below which the theory becomes four-dimensional
has been suggested by lattice simulations in \cite{Ejiri:2000fc}. The extra
dimension can even
``disappear'' when the correlation length of the four-dimensional theory grows
exponentially fast with the size of the extra dimension. This is the mechanism
of D-theory \cite{Chandrasekharan:1996ih,Schlittgen:2000xg}, which relies on
the existence of the Coulomb phase and has been recently studied on the lattice
\cite{deForcrand:2010be}. 

The most prominent of localization models in flat space\footnote{
We do not discuss here the case of warping, which modifies the metric.}
is the domain wall construction for fermions \cite{Kaplan:1992bt}, which
has been shown to be equivalent to an
orbifold construction \cite{Luscher:2000hn}. A possible localization mechanism
for gauge field has been proposed in \cite{Dvali:1996xe} and relies on a
mechanism that confines the theory in the bulk and deconfines it on a boundary
(or domain wall) of the extra dimension. This mechanism has
been studied on the lattice in \cite{Laine:2004ji}, where it was found that
indeed localization occurs but the low energy effective theory contains not just
the localized zero-modes but also higher Kaluza--Klein modes. Similarly
localization can can also be realized if a layered phase exists
\cite{Fu:1983ei,Dimopoulos:2006qz}.

The phenomenological signature of extra dimensional models depends on the
mechanism of dimensional reduction. In the compactification case a striking
evidence would be the appearance of Kaluza--Klein modes. For example,
so far the non-observation of an excitation $Z^\prime$ of the $Z$ boson at LHC
\cite{Chatrchyan:2011wq,Aad:2011xp} has put a lower bound on their mass of
about $1\,{\rm TeV}$ (assuming it has the same couplings as the Standard
Model particle). This translates into a bound on the inverse compactification
radius. In the localization case, there is a spectrum of four-dimensional
localized modes at low energies with a mass gap, which is set by the domain wall
height $M$ \footnote{On the lattice $M$ is given by the inverse lattice spacing.
}, separating the higher modes. The latter can be localized modes forming
like a Kaluza--Klein tower or bulk modes \cite{Luscher:2000hn,Laine:2004ji}. 

Recently \cite{Irges:2009bi,Irges:2009qp} the phase diagram of the
five-dimensional pure SU(2) gauge theory on the torus discretized with the
Wilson plaquette action and
anisotropic gauge couplings has been investigated in the mean-field
approximation including corrections \cite{Drouffe:1983fv,Ruhl:1982er}.
The mean-field phase diagram exhibits three phases, besides the confined and
deconfined phases there is a layered phase when when the lattice spacing
along the extra dimension $a_5$ is larger than the one in the usual four
dimensions $a_4$.
It is possible to take the continuum limit approaching a critical line (a
line of second order phase transitions\footnote{
The existence of a second order phase transition for five-dimensional
Yang-Mills theories has been hinted by the epsilon expansion
\cite{Gies:2003ic,Morris:2004mg} but has been sofar elusive in lattice Monte
Carlo simulations.}
) at the phase boundary between the deconfined and the layered phase. 
The lattice spacings go to zero by
keeping the anisotropy and the ratio of the vector to the scalar masses fixed.
In this limit the hyperplanes orthogonal to the extra dimension decouple from
each other and the theory turns four-dimensional. In the present paper we
show what happens when the full theory is simulated by Monte Carlo methods.

The paper starts with \sect{sect:definitions} containing definitions of
the lattice theory and observables used in this study. In addition to
observables (plaquettes and Polyakov loops and their susceptibilities, Binder
cumulants) signaling phase transitions, we compute the static potential in
the four-dimensional hyperplanes orthogonal to the extra dimension.
Using the latter we define
renormalized couplings from the static force and its derivative, which give us
information about dimensional reduction. In
\sect{sect:symmetric} we discuss the case of isotropic couplings (where the
lattice spacing is the same in all directions) as a starting point for the
investigations of the anisotropic case in \sect{sect:asymmetric}. 
We study the part of the phase diagram where $a_5>a_4$. We find
first order bulk phase transitions (\sect{s_asymm_bulk}), which separate the
confined from the deconfined phase,
and second order phase transitions related to breaking of the
center symmetry along one lattice dimension (\sect{s_asymm_center}).
In \sect{s_asymm_potential} we show our results for the static potential
measured at two bulk phase transition points in the confined vacuum, which
hint at dimensional reduction.
Our conclusions and outlook are
presented in \sect{sect:conclusions}. Appendix \ref{s_appa} summarizes the
formulae for the 3-loop perturbative running of the coupling derived from the
static force in the four-dimensional Yang-Mills theory, which we use to compare
with our lattice data.

\section{Definitions}
\label{sect:definitions}

We consider a Euclidean lattice in five dimensions. The dimensions are
labelled by $M=0,1,2,3,5$. The lattice size is $V=L_T\times \Ls^3\times L_5$,
where $L_T$, $\Ls$ and $L_5$ are the number of lattice points in the temporal
($M=0$), spatial ($M=1,2,3$) and extra ($M=5$) dimensions respectively.
We use Roman
capital letters $M,N,\ldots$ to label all five dimensions and Greek letters
$\mu,\nu,\ldots$ to label the four dimensions $0,1,2,3$. The lattice points
are denoted by $x=(x_0,\vec{x},x_5)$ and the gauge links by $U(x,M)$.
The Wilson lattice action for a SU(2) gauge theory is
\beq
S & = & \frac{\beta}{2} \sum_x \left[ \frac{1}{\gamma} \sum_{\mu<\nu}
  \tr\{1-U(x;\mu,\nu)\} + \gamma \sum_{\mu} \tr\{1-U(x;\mu,5)\} \right] \,,
\label{action}
\eeq
where $U(x;M,N)$ is the oriented plaquette at point $x$ in directions $M$ and 
$N$ and we use the property that the trace of SU(2) matrices is real.
The parameter $\gamma$ is the anisotropy parameter. Instead of
$(\beta,\gamma)$ we will also use the equivalent parameter pair
\beq
\beta_4 = \frac{\beta}{\gamma} \quad  &,& \quad \beta_5 = \beta\,\gamma \,.
\eeq
The isotropic lattice corresponds to $\gamma=1$.
If $\gamma\neq1$ there are two different lattice spacings, $a_4$ in the
$\mu$-dimensions and $a_5$ in the extra dimension. At the classical level
the relation $\gamma=a_4/a_5$ holds.
The relation between the lattice coupling $\beta$ and the bare dimensionful
gauge coupling $g_5$ is \cite{Ejiri:2000fc}
\beq
\beta & = & \frac{2Na_4}{g_5^2} \,,\; (N=2) \,.
\eeq
The coupling $g_5$ is an effective coupling at the cut-off scale and an
attempt to define a continuum limit $a_4\to0$ by keeping $g_5$ constant (up to
a slowly varying renormalization factor) leads to
decreasing $\beta$. Eventually the value $\beta_c$ at the phase transition
between the deconfined and confined phase is reached, meaning that the lattice
spacing cannot be further reduced. The continuum limit can therefore only be
taken at $g_5=0$, i.e. the theory is trivial \cite{Irges:2004gy}.
Triviality can also be understood by considering the 1-loop renormalization of
the effective four-dimensional coupling $g_4^2=g_5^2/(L_5a_5)$
\cite{Dienes:1998vg,Irges:2007qq}.

In the Monte Carlo simulations of \eq{action} we use the heatbath algorithm
proposed by Fabricius and Haan \cite{Fabricius:1984wp} and independently by
Kennedy and Pendleton \cite{Kennedy:1985nu} in combination with the
overrelaxation algorithm, cf. \cite{MontMuen}. One updating unit (or
iteration) is typically defined as the sequence of $\Ls/2$
overrelaxation sweeps followed by one heatbath sweep.
In order to check the ergodicity of our simulations we
have repeated some of them using Creutz's version of the heatbath algorithm
\cite{Creutz:1980zw} and a mixture of the Kennedy-Pendleton and Creutz one.
The simulation results obtained with these three variants of the algorithm
agree very well.
The statistical analysis of our simulations is usually done with
the method of \cite{Wolff:2003sm} but in some cases we use
the bootstrap method.

The observables that we measure to investigate the phase diagram
are the traces of the plaquettes, separately
for the plaquettes orthogonal to and along the extra dimension
\beq
\plaq_4 = \frac{1}{6V}\sum_x\sum_{\mu<\nu}\tr\{U(x;\mu,\nu)\} &,&
\plaq_5 = \frac{1}{4V}\sum_x\sum_{\mu}\tr\{U(x;\mu,5)\} \,,\nonumber \\
\label{plaqs}
\eeq
and their susceptibilities
\beq
\chi_{\plaq_i} & = & V\VEV{\left(\plaq_i-\VEV{\plaq_i}\right)^2} \,,\; i=4,5
\,. \label{chiplaqs}
\eeq
The plaquette observables provide signals for bulk phase transitions.
In order to study phase transitions related to the breaking of the center
symmetry we also measure Polyakov loops. For example, the Polyakov loop
along the time dimension and its susceptibility are defined by
\beq
\poly_T & = & \frac{L_T}{V}\left|
\sum_{\vec{x},x_5}\tr\prod_{x_0=0}^{(L_T-1)a_4}U(x,0)\right|
\label{polyT}
\\
\chi_{\poly_T} & = & \frac{V}{L_T}\VEV{\left(\poly_T-\VEV{\poly_T}\right)^2} \,.
\label{chipolyT}
\eeq
Polyakov loops winding in other directions are similarly defined. We will also
consider the temporal Polyakov loop without averaging over the extra
dimension, explicitely
\beq
\poly_T(x_5) & = &
\frac{1}{\Ls^3}\left|
\sum_{\vec{x}}\tr\prod_{x_0=0}^{(L_T-1)a_4}\left.U(x,0)\right|_{x_5}\right|
\label{polyTx5}
\\
\chi_{\poly_T}(x_5) & = &
\Ls^3\VEV{\left(\poly_T(x_5)-\VEV{\poly_T(x_5)}\right)^2} \,.
\label{chipolyTx5}
\eeq
It will be clear in the following why we use two different definitions
\eq{polyT} and \eq{polyTx5} of the Polyakov loop. Other
useful quantities to investigate the order of phase transitions are the Binder
cumulants \cite{Binder:1981sa}. For example, the fourth order Binder cumulant
of the temporal Polyakov loop is
\beq
B_4 & = & 1 - \frac{\VEV{\poly_T^4}}{3\VEV{\poly_T^2}^2} \,, \label{B4polyT}
\eeq
and similarly for other quantities.

Wilson loops are traces of product of links along rectangular paths (we
consider only on-axis Wilson loops). We measure Wilson loops $W(r,T)$
of size $T$ in the temporal direction and $r$ along one of the spacial
dimensions.
On anisotropic lattices there are two types of loops, depending whether the
distance $r$ is taken along the extra dimension or the other spatial
dimensions. In the measurement of the Wilson loops, the temporal links are
replaced by their one-link integrals \cite{Parisi:1983hm} and the spatial
links are replaced by the links obtained after two levels of HYP smearing
\cite{Hasenfratz:2001hp}.
The HYP smearing of the spatial links is restricted to staples extending in
the spatial directions, e.g. for a link in direction $\mu=1$ the staples are
in directions $M=2,3,5$. We use the HYP parameters
\beq
\alpha_1=0.8 \,, & \alpha_2=0.56 \,, & \alpha_3=0.24 \quad\mbox{(level 1)}\,, \\
\alpha_1=0.92 \,, & \alpha_2=0.68 \,, & \alpha_3=0.28 \quad\mbox{(level 2)}\,,
\eeq
optimized by maximizing the minimum plaquette. We find that these values do
not strongly depend on the gauge couplings. We increase the levels of smearing
up to four by using for level 3 and 4 the same parameters as for level 2. We
checked numerically that the average and minimum plaquette increase with the
number of smearing levels since our parameters $\alpha_1$ are larger than
the perturbative bound on the APE smearing parameter
$\alpha\le0.75$ \cite{Bernard:1999kc}. From the Wilson loops we compute
the effective masses
\beq
a_4m_{\rm eff}(r,t+a_4/2) & = & 
\ln\left(\frac{\VEV{W(r,t)}}{\VEV{W(r,t+a_4)}}\right) 
\;\stackrel{t\to\infty}{\sim}\; a_4V(r)
\eeq
and extract the static potential $V(r)$ from a plateau average of the effective
masses at large enough $t$, cf. \cite{Donnellan:2010mx}. In order to
determine where to start the plateau average for each value of $r$ we perform
a fit
\beq
m_{\rm eff}(r,t+a_4/2) & = &
 E + b\,{\rm e}^{-(t+\frac{a_4}{2})\Delta} \label{fiteff}
\eeq
with fit parameters $E$, $b$ and $\Delta$. The plateau average starts at the
earliest time $t_0$ when the exponential correction in \eq{fiteff} is smaller
than 1/4 of the the statistical error of the effective mass.
The plateau ends before the time $t$ when
either the difference of the effective mass at time $t$ to the one at $t_0$
is larger than the error of the effective mass at $t_0$
or the error of the effective mass at time $t$ is larger than twice the error
of the one at $t_0$.

In our analysis we focus on the static potential along spatial
directions orthogonal to the extra dimension and we call this potential
$V(r)$. We define the static force through the symmetric derivative of the
potential
\beq
F(r-a_4/2) & = & \{V(r)-V(r-a_4)\}/a_4 \,. 
\eeq
The scale $r_0/a_4$ \cite{Sommer:1993ce} is determined through the numerical
solution of
\beq
\left. r^2\,F(r)\right|_{r=r_0} & = & 1.65 \,, \label{r0}
\eeq
which we find using a 2-point interpolation of the force
$F(r)=f_0+f_2/r^2$. A comparison with a 3-point interpolation adding a
$f_4/r^4$ term allows to estimate the systematic error.
From the static force, a running coupling can be defined (in the so called
qq-scheme)
\beq
\aqq(1/r) & = & \frac{1}{\CF}\,r^2\,F(r) \,, \label{aqq}
\eeq
where $\CF=3/4$ for gauge group SU(2).
The perturbative expansion up to three loops of $\aqq$ in four-dimensional
Yang--Mills theory is summarized in Appendix \ref{s_appa}.
By taking a further derivative we compute the slope \cite{Luscher:2002qv}
\beq
c(r) & = & \frac{1}{2}\,r^3\,F^\prime(r) \,, \label{ccoeff}
\eeq
which also defines a running coupling (at small distances). On the lattice we
compute $c(r)$ through
\beq
c(r) & = & \frac{1}{2}\,r^3\,\{V(r+a_4)+V(r-a_4)-2V(r)\}/a_4^2 \,. 
\eeq
At large
distances, the slope can be compared to the result of effective bosonic string
theory \cite{Luscher:1980fr,Luscher:1980ac}, which yields the asymptotic value
\beq
c(\infty) & = & -\frac{(D-2)\pi}{24} \,,
\eeq
where $D$ is the number of space-time dimensions. Thus $c$ can serve to
determine the number of dimensions.
We remark that due to the factor $r^3$ in its definition, the noise-to-signal
ratio of $c(r)$ rapidly deteriorates when $r$ grows.

Since most of our large volume simulations are done at $\gamma<1$ in the
confined phase, it turns out that the
static potential along the extra dimension $V_5(r)$ is difficult to
extract. The reason is that the values of $a_4V_5(r)$ we measure at $r=2a_5$
are typically 1 or larger and the dimensionless string tension (the
coefficient of the term in $V_5(r)$ which is linear in $r$) $a_4a_5\sigma_5$
contains the large lattice spacing $a_5$.

\section{Isotropic couplings \label{sect:symmetric}}

The aim of this section is to review known results for and highlight features
of the isotropic case ($\gamma=1$)
that will be of help in the understanding of the anisotropic ($\gamma\neq1$)
case.
We will start with the description of the system at zero temperature in the 
``infinite'' volume limit.
Afterwards we will analyze how any kind of finite size affects the results.
\begin{figure}[t]
  \begin{center}
   \includegraphics*[angle=-90,width=.48\textwidth]{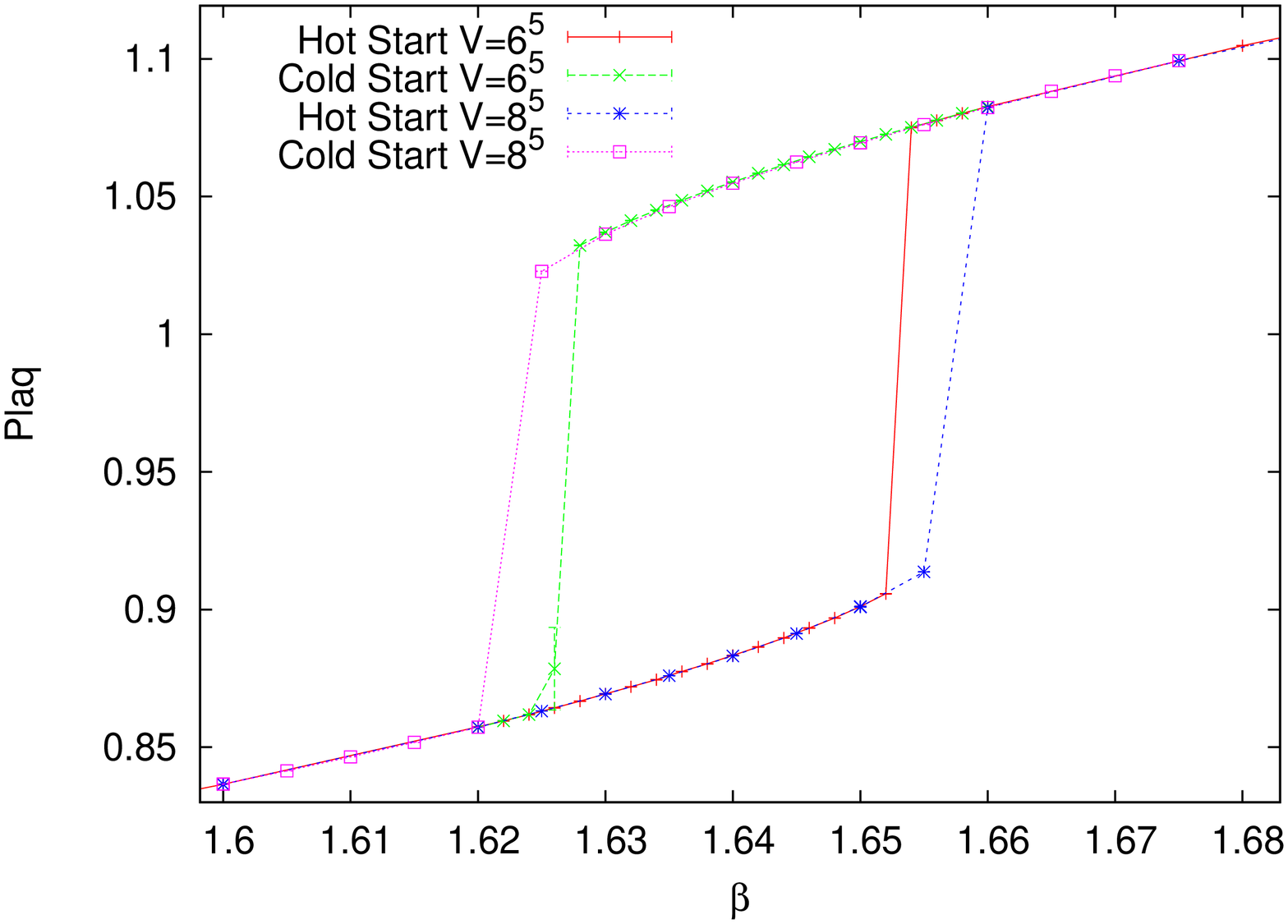}
   \includegraphics*[angle=-90,width=.48\textwidth]{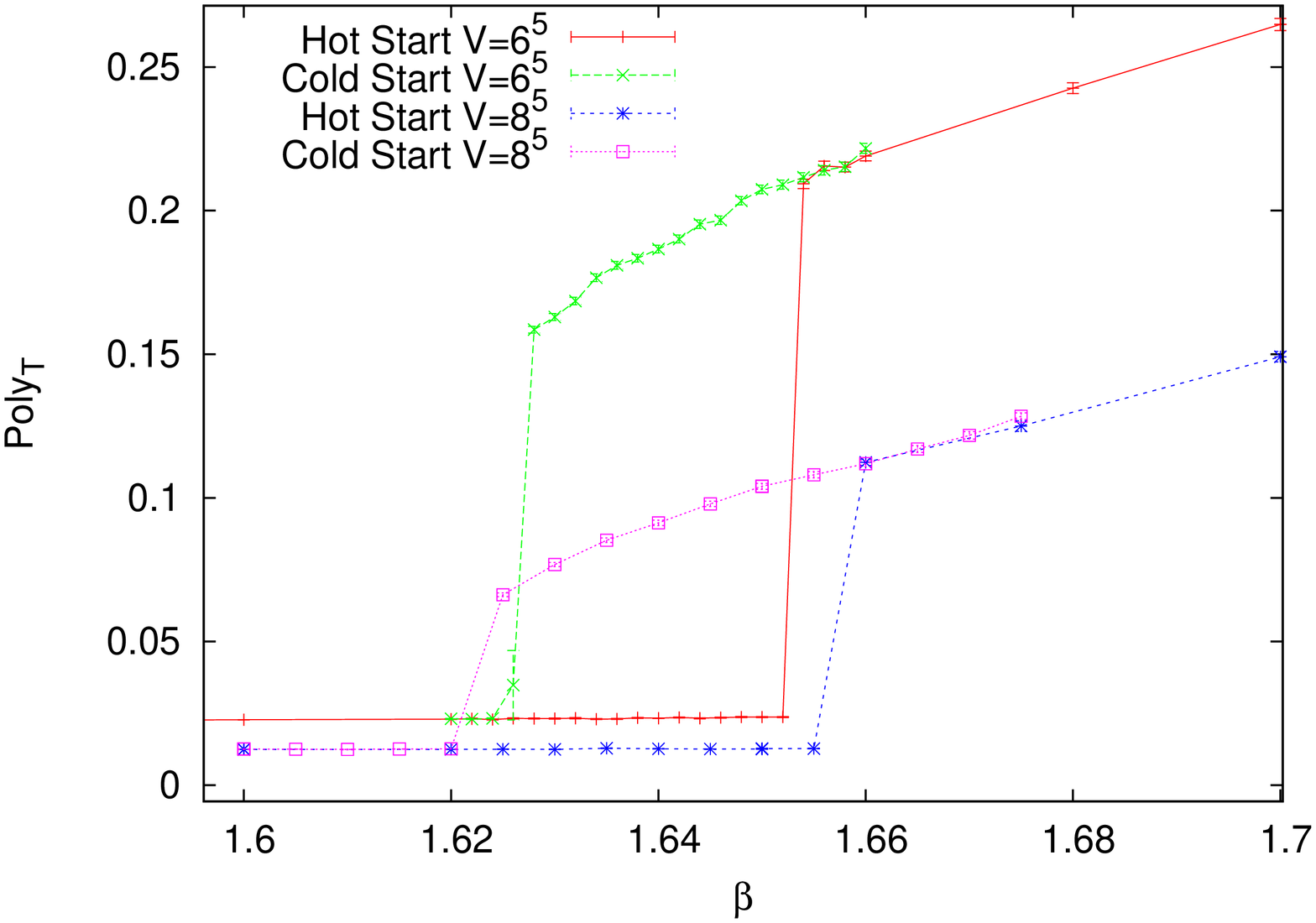}
\end{center}
  \caption{\small Plaquette and Polyakov expectation value as function of
    $\beta$ on isotropic lattices.}
  \label{fig:gamma1_1}
\end{figure}

The phase diagram of the five-dimensional isotropic SU(2) lattice gauge theory
is well known since the early work of Creutz \cite{Creutz:1979dw}.
In the zero temperature setting and at infinite volume (both spatial and
in the extra-dimension) the phase diagram of the model is characterized by 
the presence of a bulk first order phase transition at a critical value 
of the lattice coupling $\beta=\beta_c\approx1.64$.
The bulk phase transition is 
characterized by the appearance of a strong hysteresis region in the expectation
value of the plaquette, see \fig{fig:gamma1_1}, that survives in the infinite
volume limit, signal of the presence of latent heat at the transition point.
The two regions separated by $\beta_c$ have a different
behavior of the Wilson loop, with a string tension (the coefficient $\sigma$
of the linear term $\sigma\,r$ in the static potential $V(r)$) different from
zero in the confined phase for $\beta<\beta_c$ and a string tension equal to
zero in the deconfined phase above $\beta_c$.
On a finite (but large) lattice this coincides with a distribution of the 
Polyakov loop (for all directions and not taking the absolute value) being
always peaked in zero in the first case, while always showing a double peak
structure in the second case.

In \fig{fig:gamma1_1} we report an example of the typical plots showing 
the behavior at the phase transition both for the expectation value of the 
plaquette and of the absolute value of the temporal Polyakov loop
$\poly_T$ defined in \eq{polyT}.
``Hot Start'' or ``Cold Start'' means that the simulation starts from an
initial configuration with random gauge links or gauge links set to the
identity, respectively. We perform $10^4$ measurements separated by 10
updating iterations.

\subsection{One ``small'' dimension}

Now we investigate the effects of the finite lattice size 
on our observables and how they affect the bulk phase transition.
As it is well known there is no sign of first order bulk phase transition in 
four-dimensional SU(2) Yang--Mills theory, instead a sharp crossover separates
the weak and the strong coupling regime. It is then
legitimate to ask which is the fate of the five-dimensional bulk phase
transition when we consider the five-dimensional system compactified along
one ``small'' direction.
\begin{figure}[t]
  \begin{center}
   \includegraphics*[angle=-90,width=.48\textwidth]{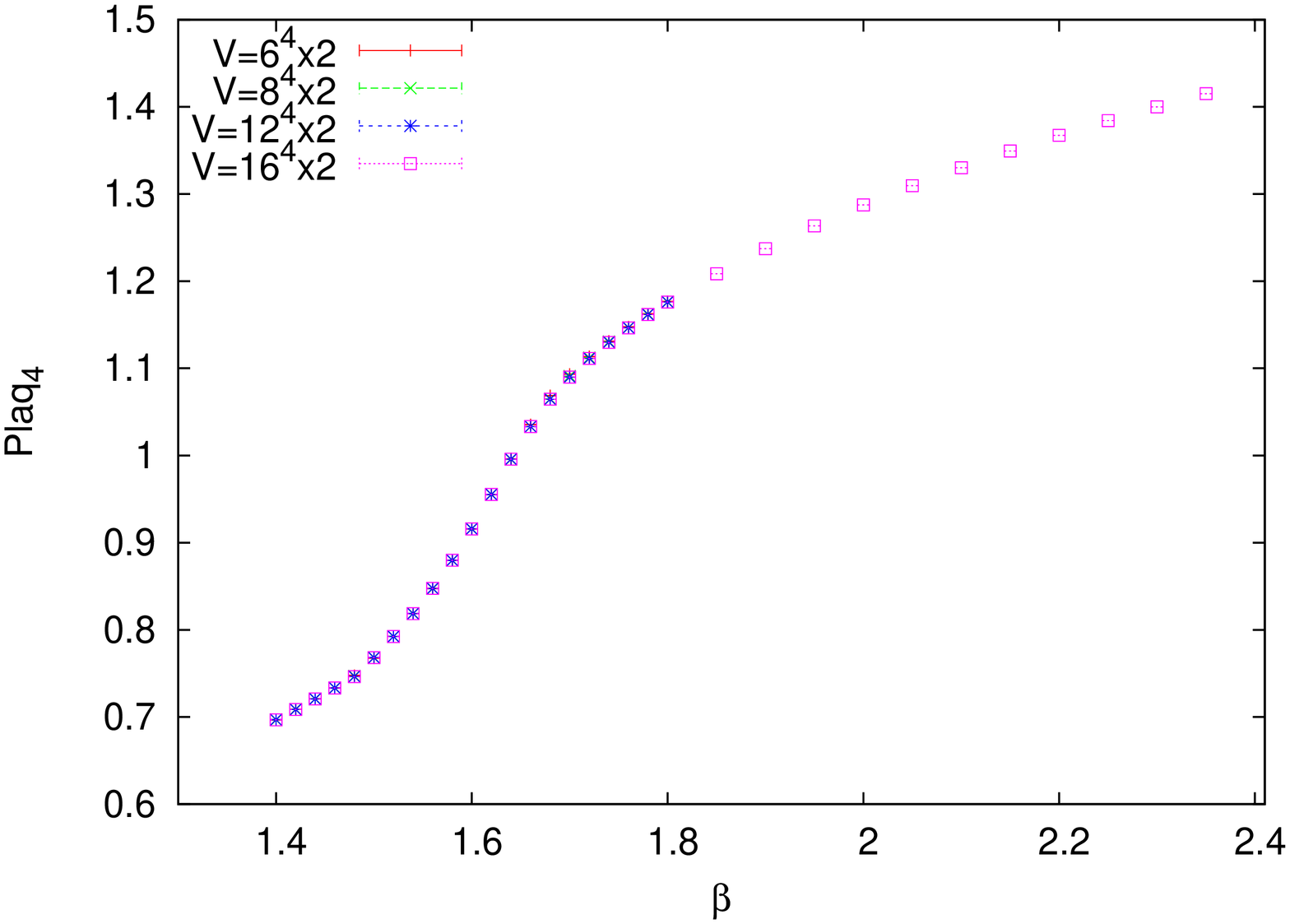}
   \includegraphics*[angle=-90,width=.48\textwidth]{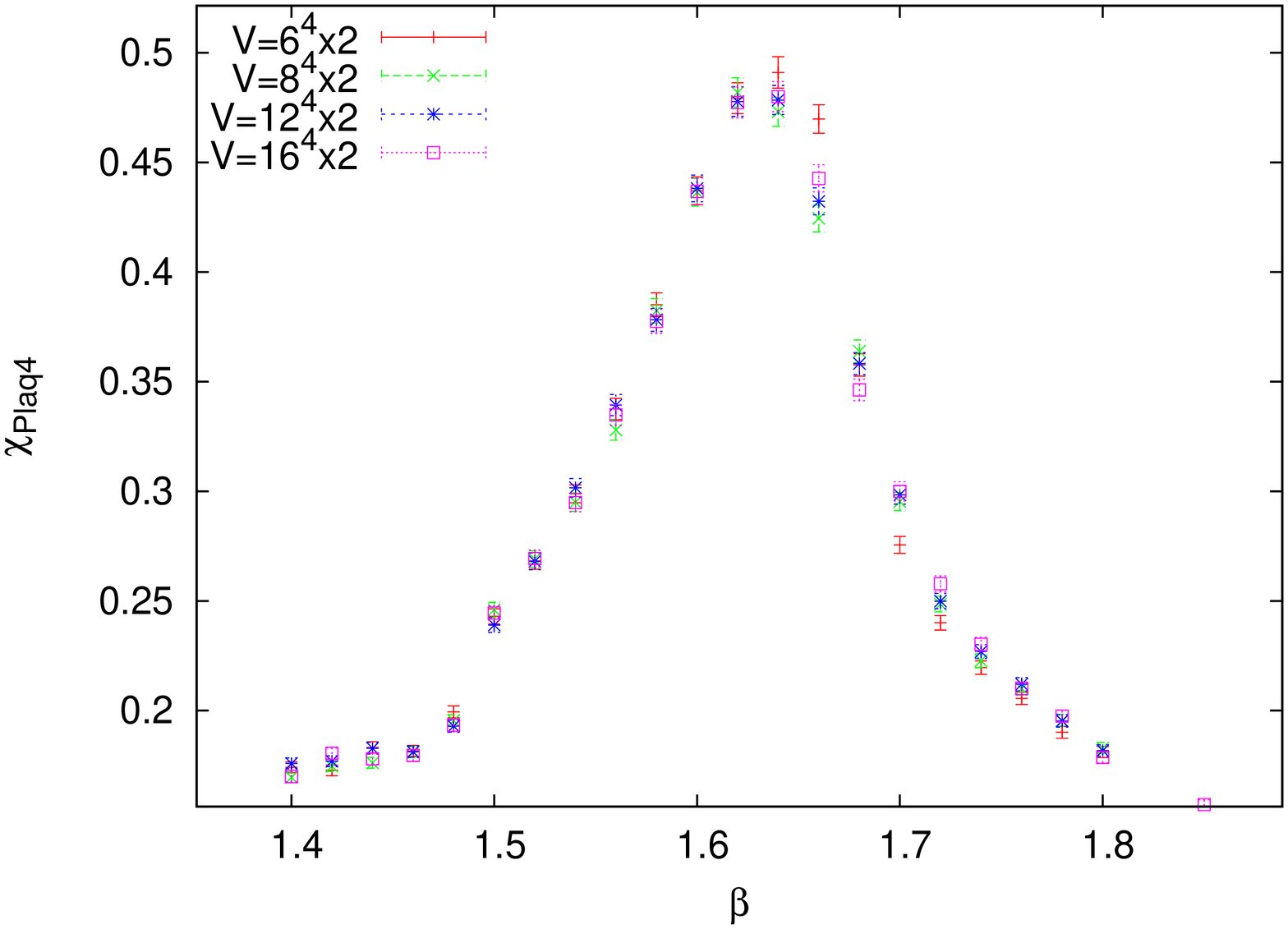}
\end{center}
  \caption{\small Expectation values of the plaquette $\plaq_4$ and its
    susceptibility $\chi_{\plaq_4}$ as function of $\beta$ on
    lattices with $L_5=2$ and different values of $L_T=\Ls=6,\ldots,16$.}
  \label{fig:gamma1_2}
\end{figure}

If we consider a geometry that has one small direction
while all the other directions are sent to infinity, 
namely $L_5\ll L_T=\Ls$, we notice that the 
bulk phase transition at $\beta=1.64$ ``evaporates'' and lets a smooth
cross-over appear, cf. the early work of \cite{Lang:1986kq}.
This is shown by the behavior of the four-dimensional plaquette $\plaq_4$
\eq{plaqs} and of its susceptibility $\chi_{\plaq_4}$ \eq{chiplaqs} 
in \fig{fig:gamma1_2}, for $L_5=2$ and different choices of the
four-dimensional lattice size $L_T=\Ls=6,\ldots,16$.
There is no dependence nor on the initial condition
(hot or cold start) of the simulation neither on the lattice volume.
The cross-over region is close to $\beta=1.48$ and will be investigated below.
The susceptibility of the plaquette shows a peak at $\beta$ values reminiscent
of the bulk phase transition but whose magnitude is independent on the volume. 
We extend the simulations for
our largest $2\times16^4$ lattice up to $\beta=2.4$ but we do not see any sign
of the cross-over of the four-dimensional SU(2) theory, which is at
$\beta\approx2.3$ \cite{Bonati:2009pf}.

Our simulations hint at the existence of a minimal lattice size\footnote{
We are exposing the dependence on $\gamma$ in order
to extend later the definition of the minimal size to the case $\gamma\neq1$.}
\beq
 2 < \Lmin(\gamma=1) < 6 \,, \label{lming1}
\eeq
for which if one or more of the lattice dimensions are smaller
then $\Lmin(\gamma=1)$ we are
unable to detect any sign of the bulk phase transition.
\begin{table}
\begin{center}
\begin{tabular}{c|c|c|c|c}
Volume & $\beta_c$ ($\poly_T$) & $\beta_c$ ($\poly_T(1)$) &
$\chi_c$ ($\poly_T$) & $\chi_c$ ($\poly_T(1)$)\\
\hline
$2\times4^4$      & 1.50061(19)  & 1.50369(23)
                  & 11.86(4)     & 3.842(15)   \\
$2\times6^4$      & 1.49097(13)  & 1.49249(14) 
                  & 30.00(13)    & 6.571(19)   \\
$2\times8^4$      & 1.48748(11)  & 1.48839(10) 
                  & 56.9(4)      & 9.44(5)     \\
$2\times10^4$     & 1.48570(10)  & 1.48621(12) 
                  & 93.1(8)      & 12.19(9)    \\
$2\times16^4$     & 1.483952(16) & 1.484170(18) 
                  & 257.8(1.6)   & 20.88(10)   \\
\hline
$2\times\infty^4$ & 1.48280(1)   & 1.48280(3)  & - & -
\end{tabular}
\end{center}
\caption{\small For the center breaking phase transition at $\gamma=1$,
  $L_T=2$, we list
  the critical values of the coupling $\beta_{c}$ and the
  susceptibility of the temporal Polyakov loop $\chi_c$
  as function of the volume. We use
  two definitions of the Polyakov line, \eq{polyT} and
  \eq{polyTx5}.}
\label{tab:1_sym}
\end{table}
 
Apart from the study of the plaquette, other interesting signals come from
the study of the expectation value of the Polyakov loops and their
susceptibility in the geometry with one small direction.
It is interesting to notice that together with the disappearance of the bulk 
phase transition also the double peak structure of the Polyakov line 
distribution along the large directions disappears.
On the other side we expect, and find, a proper center breaking structure 
along the small direction, which we choose for this study to be the temporal
one. In order to avoid the appearance of the bulk phase transition we
always request that $L_T<\Lmin(\gamma=1)$.
Notwithstanding the fact that we need to fix $L_T$ to
a finite value, we can send the orthogonal space size $L=\Ls=L_5$ to infinity
and obtain a system that undergoes a proper phase transition.
We performed a study of the finite size scaling properties of this breaking 
by investigating both the distribution of the temporal Polyakov loop and its 
Binder cumulant (as discussed in \cite{deForcrand:2010be}) as function of the
orthogonal lattice size, while keeping $L_T=2$ fixed.
\begin{figure}[t]
  \begin{center}
   \includegraphics*[angle=-90,width=.60\textwidth]{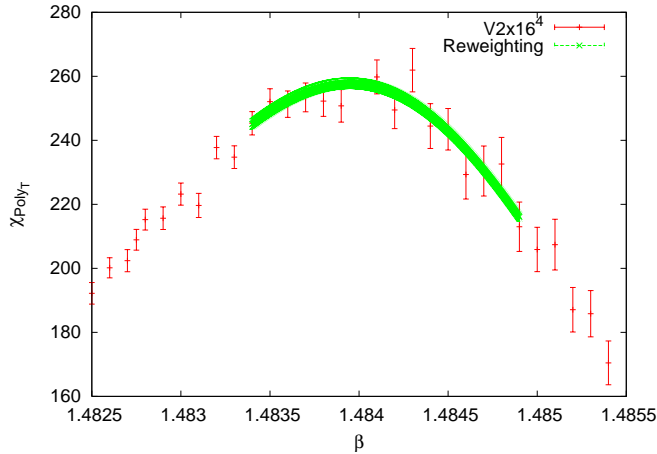}
\end{center}
  \caption{\small
The susceptibility of $\poly_T$ for the volume $2\times16^4$. Comparison of
the directly simulated data set (plusses) and the reweighted data (crosses).
}
  \label{fig:gamma1_6}
\end{figure}

We did a scan in $\beta$ for the different choices of the lattice
volume and we used a reweighting technique to obtain a more dense scan. The
technique is the multi-histogram reweighting method
\cite{m.01:mcm_statphys}. Its main idea is to find iteratively
the free energy of the system as function of the couplings using all the
simulations performed in the region of interest. Once the procedure has
converged, we have access via the free energy to all the physical quantities
that can be normally extracted in a Monte Carlo simulation, for each value of
the coupling constants in the region. Our implementation is able to handle
simultaneous reweighting in both the couplings $\beta_4$ and $\beta_5$, even
though in this section we are interested in the $\beta_4=\beta_5$ case.
\begin{figure}[t]
  \begin{center}
   \includegraphics*[angle=-90,width=.48\textwidth]{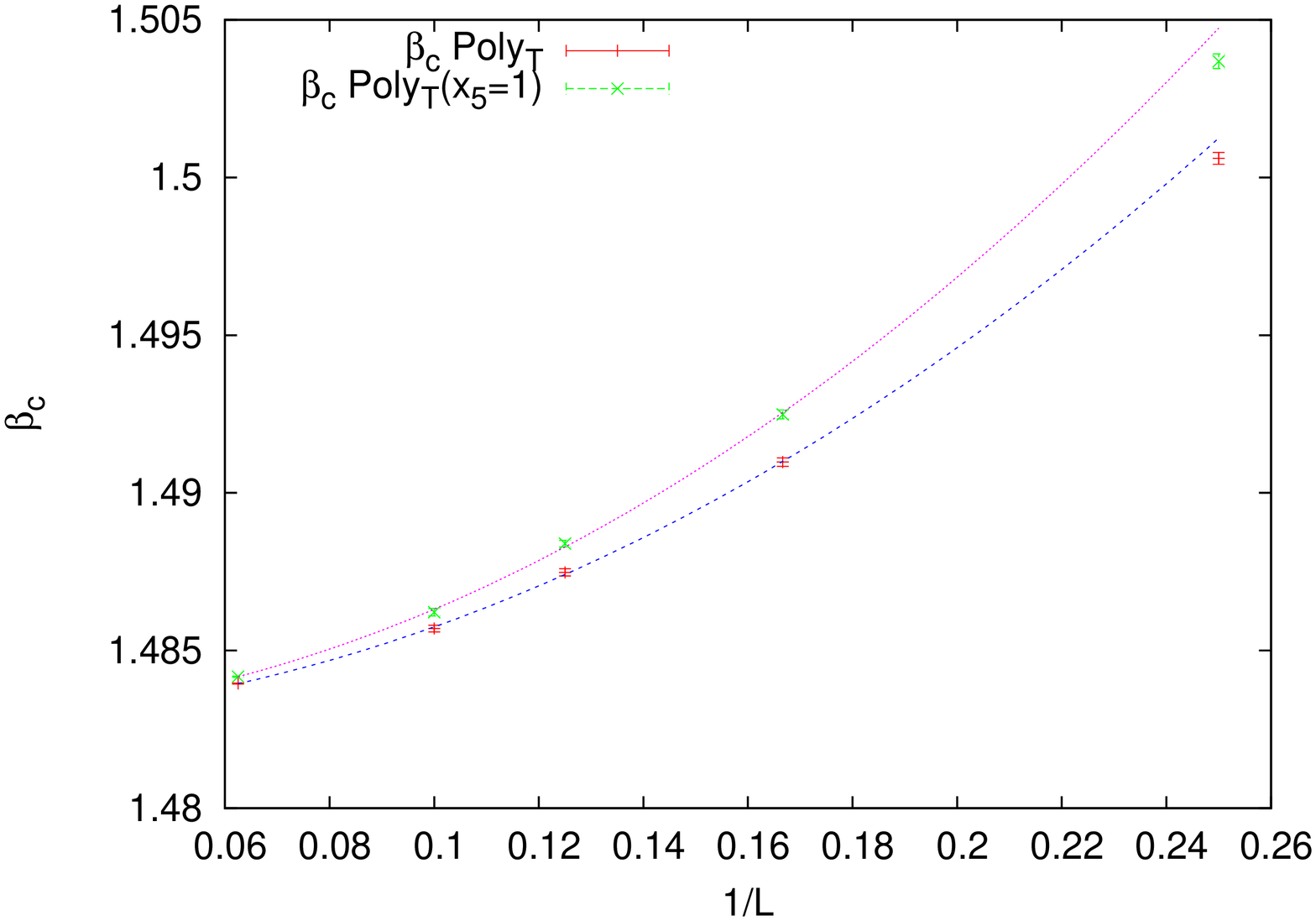}
   \includegraphics*[angle=-90,width=.48\textwidth]{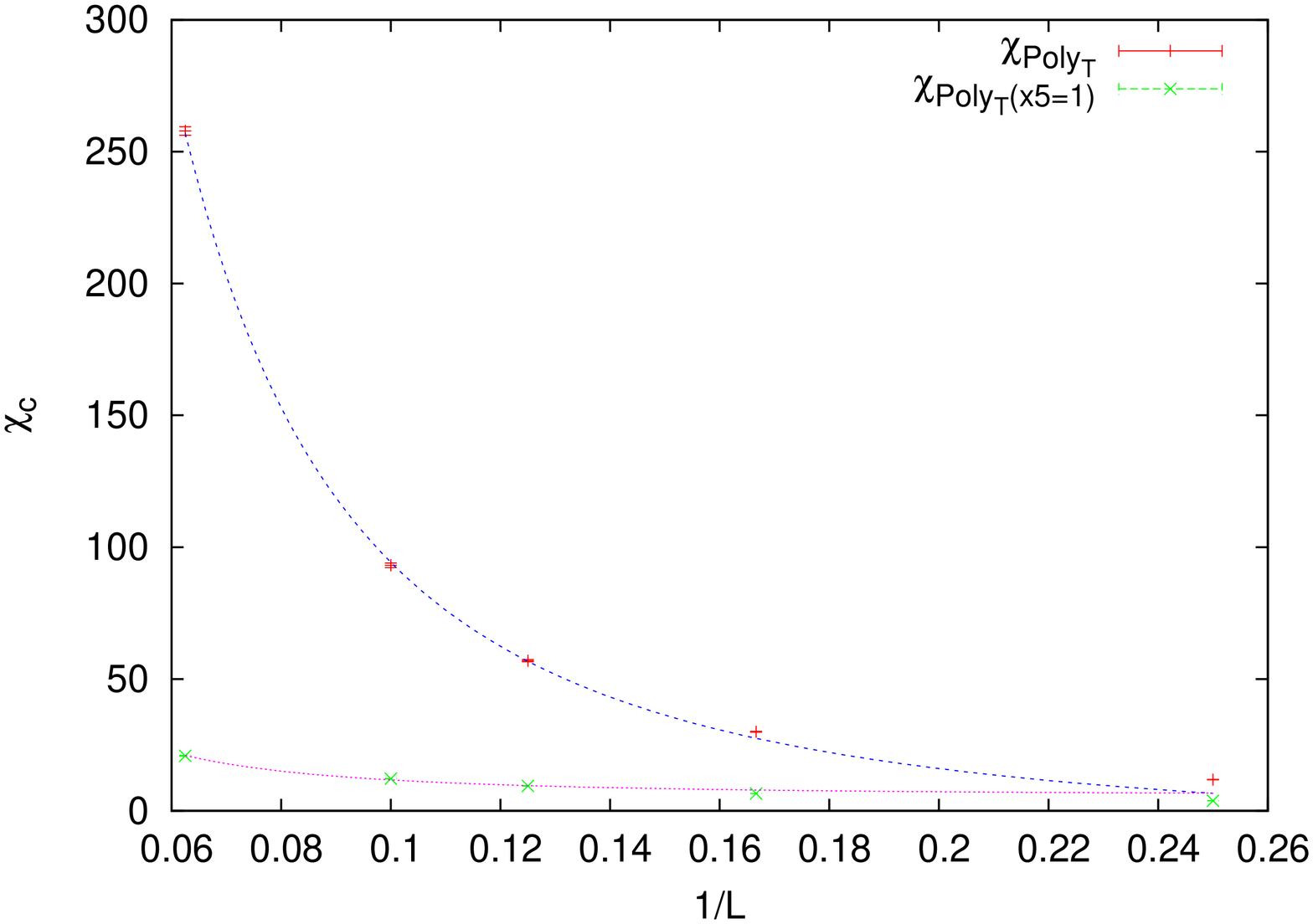}
\end{center}
  \caption{\small Finite size scaling analysis for the transition at
    $\gamma=1$, $L_T=2$ based on \tab{tab:1_sym}.
    We use two definitions of the temporal Polyakov loop,
    \eq{polyT} (plusses) and \eq{polyTx5} (crosses). 
    The lines are fits to the data using the critical exponents of the
    four-dimensional Ising model.}
  \label{fig:gamma1_3}
\end{figure}
\begin{figure}[th!]
  \begin{center}
   \includegraphics*[angle=-90,width=.60\textwidth]{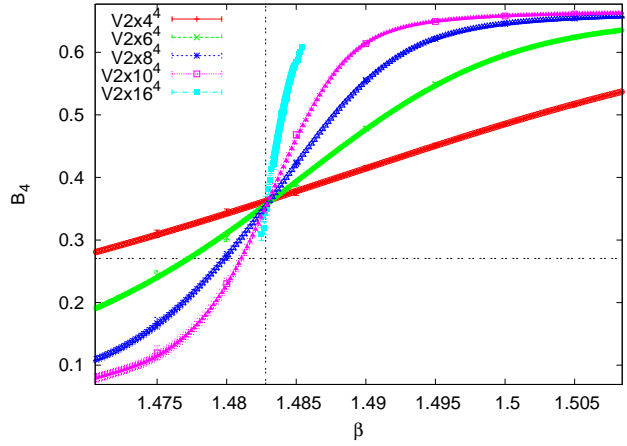}
\end{center}
  \caption{\small Finite size scaling analysis for the transition at
    $\gamma=1$, $L_T=2$: the Binder
    cumulant $B_4$ of $\poly_T$ as a function of $\beta$ for several values of
    $L=\Ls=L_5$.}
  \label{fig:gamma1_4}
\end{figure}

In order to determine the error of our estimates we use the bootstrap method
and do a reweighting analysis over each bootstrap sample.
The underlying idea is to consider the estimates coming from each bootstrap
sample as independent measurements and to use the spread of these
measurements to define the error of the average of the estimates.
In \fig{fig:gamma1_6} we clearly see that our reweighted estimates for the
susceptibility $\chi_{\poly_T}$ (crosses, they appear like a band in the
figure) \eq{chipolyT} are more precise than the measurements themselves
(plusses). This is however not surprising, since the reweighting technique
uses for each estimate the statistical information coming from all the simulated
points.

The critical coupling $\beta_c$ is
defined as the coupling at which the susceptibility $\chi_{\poly_T}$
has its maximum $\chi_c\equiv\chi_{\poly_T}(\beta_c)$.
The results for $\beta_c(L)$ and $\chi_c(L)$ as a function of the lattice
size $L=\Ls=L_5$ are listed in \tab{tab:1_sym}. We plot them in
\fig{fig:gamma1_3}, where the plusses refer to the Polyakov line averaged
along the extra dimension, \eq{polyT} and \eq{chipolyT}, and the crosses refer
to the Polyakov line evaluated in the
first slice along the extra dimension, \eq{polyTx5} and \eq{chipolyTx5}. 
In order to define the error of the derived observables $\beta_c(L)$ and
$\chi_c(L)$ at the critical point we used again the bootstrap procedure.
We studied both the bootstrap analysis of the reweighted susceptibility
and the distribution of the
critical value fitted on each bootstrap sample.
The first analysis is used only
to define a fitting range for the second one.
The critical behavior of a second order phase transition is
\beq
\chi_c(L)\,\equiv\,\chi_{\poly_T}(\beta_c(L)) & \sim & L^{\gamma/\nu}
\,,\label{2ndpt_1} \\
|\beta_c(L)-\beta_c(L=\infty)| & \sim & L^{-1/\nu} \,. \label{2ndpt_2}
\eeq
We do not have enough data to predict in a reliable way the
critical exponents. But what we can clearly state is that all our predictions
are totally compatible with a phase transition of second order and the scaling
behavior dictated by the critical exponents of the four-dimensional Ising
model, where $\gamma$ and $\nu$ are respectively $1$ and $1/2$, as expected
from \cite{Svetitsky:1982gs}. The lines in
\fig{fig:gamma1_3} are fits to our data using the critical exponent of the
four-dimensional Ising model. For $\beta_c$ we fit the data for $L>4$ and the
fits work well. For $\chi_c$ we fit $L>6$ but there are still some non-scaling
features which means that we would need bigger volumes.
We estimate $\beta_c(L=\infty)=1.48280(1)$. There is perfect agreement between
the determination of $\beta_c$ using both definitions of $\poly_T$,
see \tab{tab:1_sym}. We emphasize that the two definitions
of the Polyakov line must (and indeed they do) lead to the same value for the
critical coupling in infinite volume and the same value for the critical
exponents. We notice that the deviations from scaling are larger for the
definition without the average along the extra dimension and the
autocorrelation times are bigger but still under control.
\begin{figure}[t]
  \begin{center}
   \includegraphics*[angle=0,width=.48\textwidth]{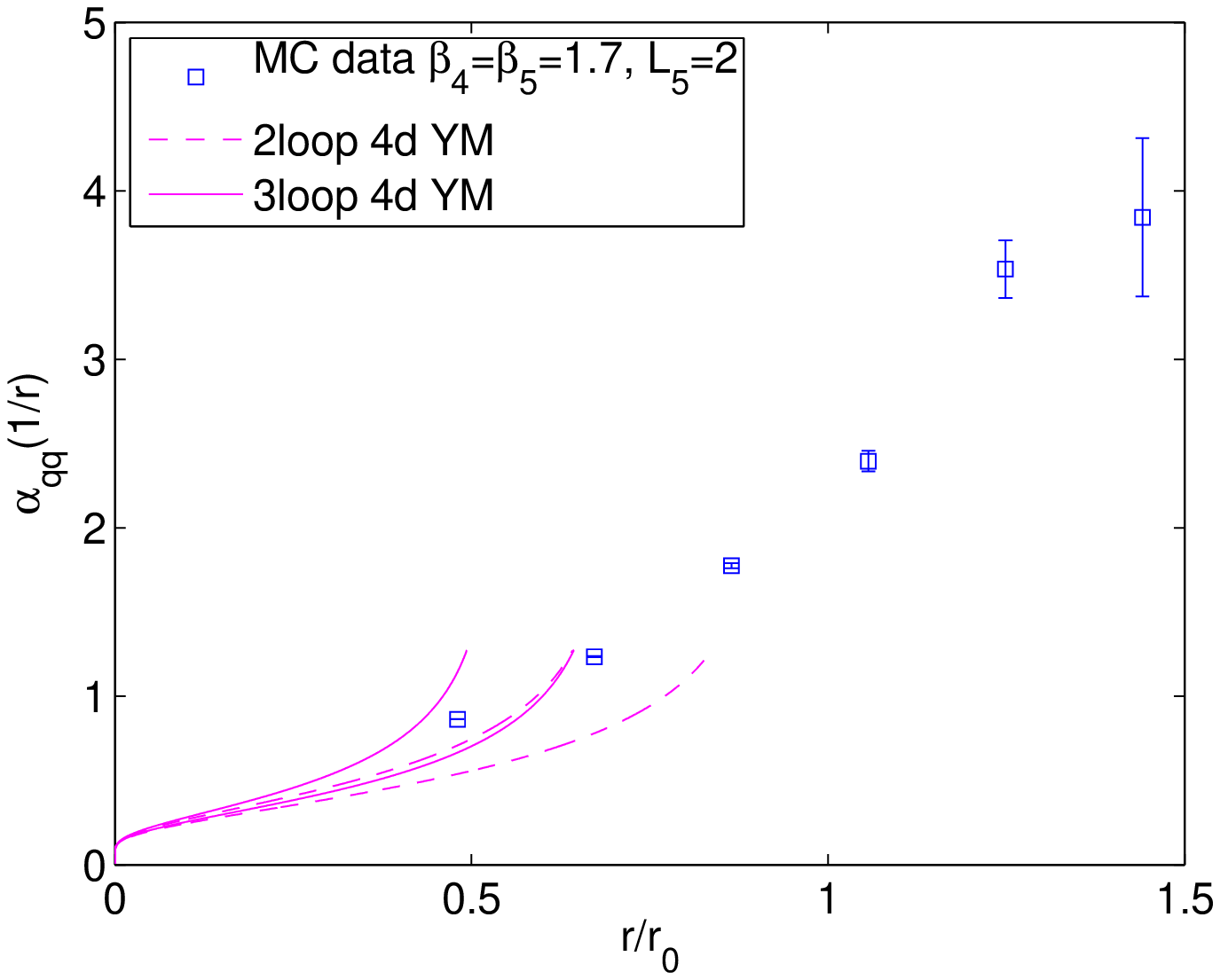}
   \includegraphics*[angle=0,width=.48\textwidth]{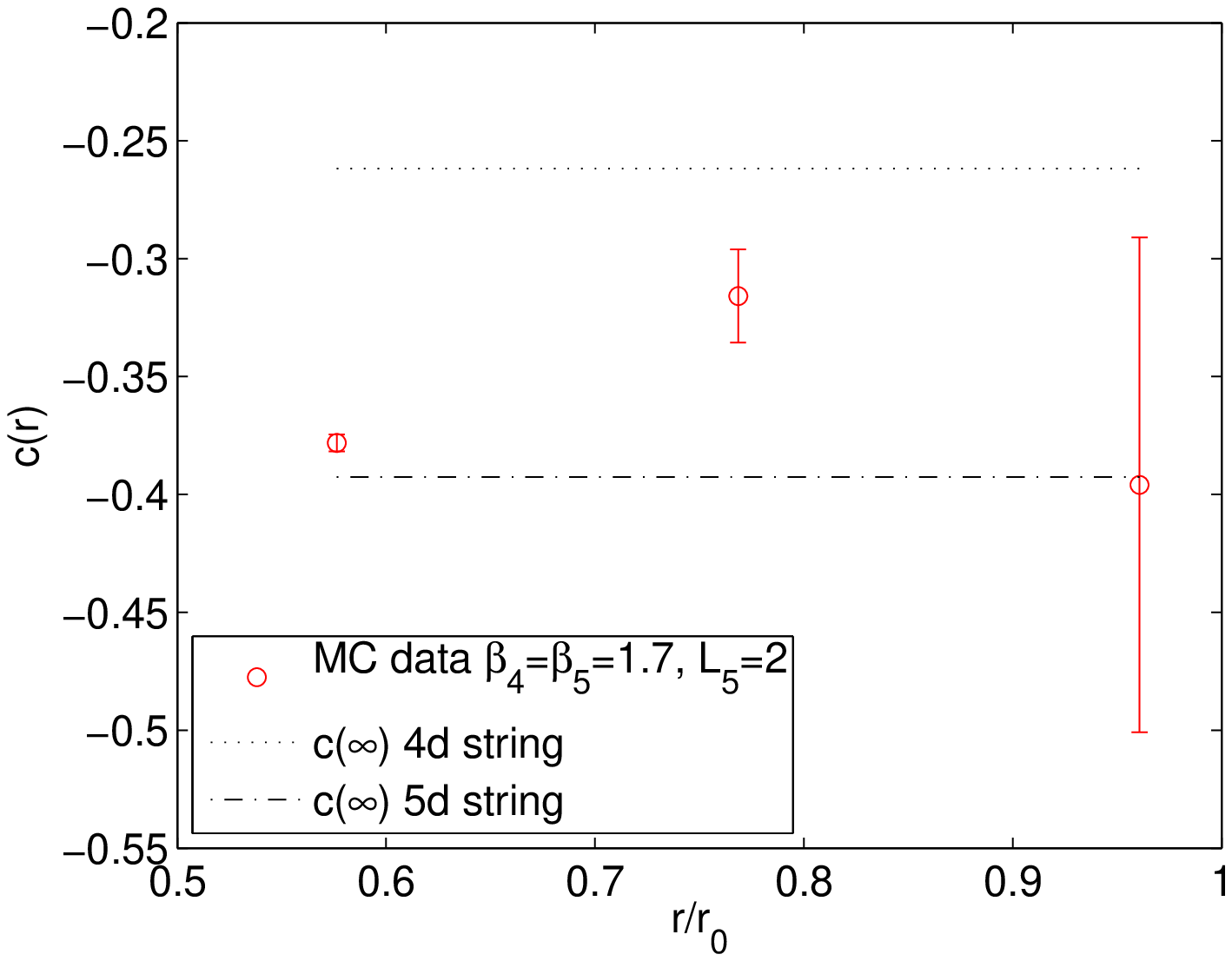}
\end{center}
  \caption{\small The coupling $\aqq$ and the slope $c$ on a isotropic lattice with
    $L_5=2$ at $\beta=1.7$.}
  \label{fig:gamma1_5}
\end{figure}

In \fig{fig:gamma1_4} we show the Binder cumulant $B_4$ of the temporal
Polyakov loop \eq{B4polyT} (here we use only the definition of $\poly_T$ in
\eq{polyT}). We plot the directly simulated (symbols in the legend) and the
reweighted data together. 
The Binder cumulants tend to intersect at the value
$\beta_c(L=\infty)$ marked by a vertical line.
Looking at \fig{fig:gamma1_4} on the right hand side of
the intersection ``point'' the smallest volume ($L=4$) corresponds to the
lowermost data points and the largest volume ($L=16$) to the uppermost
data points.
The value of the intersection is for our small lattice 
sizes $L$ still far from the analytic estimate for the universality class of
the four-dimensional Ising model $B_4^c\approx0.27$ (horizontal
line in \fig{fig:gamma1_4}) \cite{deForcrand:2010be}, indicating again that
larger values of $L$ would be necessary for a complete analysis.
As discussed in \cite{deForcrand:2010be} the values of $B_4$ computed at
$\beta_c(L=\infty)$ approach $B_4^c$ with corrections which slowly decrease
like $1/\ln(L)$. For our volumes an extrapolation of $B_4$ at $L=\infty$ is
not feasible. Our strongest evidence for the universality class of the
four-dimensional Ising model comes from the scaling analysis
presented in \fig{fig:gamma1_3}.

We have computed the static potential $V(r)$ on 
$32^4\times2$ lattices at $\beta=1.7$. 
We ran 4 replica for a total of $4\times10^4$ measurements. Each
replicum used 64 cores of the supercomputer Cheops of the University of
Cologne and consists of $10^4$ update iterations (each iteration does one
heatbath and 16 overrelaxation sweeps) for thermalization and
$10^4$ measurements of the Wilson loops separated by 10 update iterations.
We take two levels of HYP smearing for the spatial links of the Wilson loops.
We measure the potential starting from distance $r=2a$ and the force from
$r=2.5a$.
For the fit of the effective masses in \eq{fiteff} we use the range
$t=2,3,\ldots,6$.
We obtain the scale $r_0/a=5.21(7)$ with an
integrated autocorrelation time $\taui(r_0)=0.5$.
The potential can be fitted for $r>r_0$ by the form predicted from
the effective bosonic string theory and we estimate the string
tension to be $\sigma\,r_0^2=1.6(3)$.
The left plot in \fig{fig:gamma1_5} shows our data for the coupling 
$\aqq(1/r)$ defined in \eq{aqq} as a function of $r/r_0$.
For comparison we plot the 2-loop
(dashed) and 3-loop (solid) perturbative curves for the SU(2) Yang--Mills theory
in four dimensions, as explained in Appendix \ref{s_appa}. The
data at our smallest distance agree with the 3-loop perturbative curve. 
The results for the slope $c(r)$ defined in \eq{ccoeff} are shown in the right
plot of \fig{fig:gamma1_5}. There is a trend towards
the value $c(\infty)=-\pi/12$ predicted by the effective bosonic string in four
dimension (marked by the dotted line), but the statistical errors are already
too large at distance $r=r_0$.

\section{Anisotropic couplings \label{sect:asymmetric}}

In order to outline the phase diagram of the system at $\gamma\neq1$ we 
start from the already known results for $\gamma=1$ and follow the evolution
of the various ``critical points'' as function of $\beta_4$ and $\beta_5$ for
the different lattice geometries that we presented.
The critical lines can be cataloged in the following groups:
\begin{itemize}
\item 
bulk phase transitions.\\
This kind of transition seems to be always present if the geometry of the
lattice is large enough, i.e. in the ``infinite volume'' and zero temperature 
limit. More precisely for any fixed $\gamma$ we find a window of
values of $\beta_4$ and $\beta_5$ in which we detect a bulk phase transition 
signaled by an hysteresis effect in the quantities $\pq{4}$ and $\pq{5}$.
The hysteresis is seen provided the volume is large enough\footnote{
We extend the definition of the minimal size \eq{lming1} in order to
keep track of the different lattice spacings $a_4$ and $a_5$.}:
\beq
L_5>\L5min(\gamma) & \mbox{and} & \min\{\Ls\,,\,L_T\}>\Lsmin(\gamma) \,. 
\label{lmin}
\eeq
\item Center breaking phase transitions in the temporal direction (or
  analogously in any other space direction other than the extra dimension).\\
This kind of transition can be obtained by considering a small
temporal direction $L_T<\Lsmin(\gamma)$ and the other directions larger than
their respective minimal sizes. The order parameter is the
Polyakov loop $\poly_T$.
\item Center breaking phase transitions in the extra direction.\\
This kind of transition is obtained with a small extra dimension
$L_5<\L5min(\gamma)$ while the other dimensions are larger than
$\Lsmin(\gamma)$. The order parameter is the Polyakov loop winding along the
$M=5$ direction.
\end{itemize}

\subsection{Bulk phase transitions \label{s_asymm_bulk}}
\begin{table}
\begin{center}
\begin{tabular}{c|c|c|c}
$\beta_{4c}$ & $\beta_{5c}$ & $\Ls$ & $r_0/a_4$ \\
\hline
1.865    & 1.34        & 10  &\\
1.955    & 1.24        & 10  &\\     
2.33     & 0.937       & 14  & 4.774(44)\\
2.5      & 0.8697      & 20  & 8.46(11)
\end{tabular}
\end{center}
\caption{\small Critical couplings for the bulk phase transition and
  estimates $\Ls$ for the minimal lattice size $\Lsmin$ required to be in large
  four-dimensional volume and see the first order transition. 
  Where we determined it, we list the values of the scale $r_0/a_4$.}
\label{tab:1_asymbulk}
\end{table}
\begin{figure}[t]
  \begin{center}
   \includegraphics*[angle=0,width=.70\textwidth]{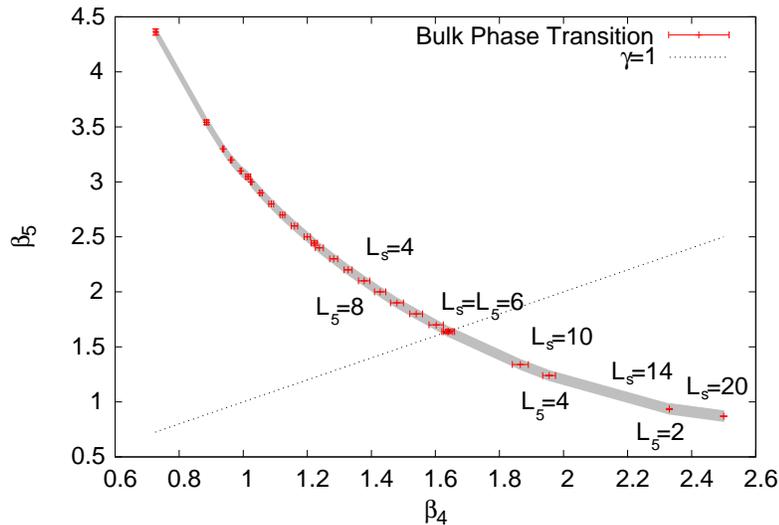}
\end{center}
  \caption{\small
Bulk phase transitions (plusses, the grey band is to guide the eye)
of the five-dimensional SU(2) gauge theory using the anisotropic
Wilson plaquette action. We indicate estimates $\Ls$ and $L_5$
for the minimal values $\Lsmin$ and $\L5min$, see \eq{lmin}, required to see
that the transitions in this plot are first order.}
  \label{fig:asymm_bulk}
\end{figure}

\fig{fig:asymm_bulk} shows the behavior of the bulk phase transition in the
$(\beta_4,\beta_5)$ parameter plane. The points (plusses) have been determined
by simulations and their ``errors'' are the width of the hysteresis on the
largest lattices we have simulated and are therefore only indicative.
The phase transition is everywhere first
order and we draw a grey band through the points to guide the eye.
We put values for $\Ls$ (that
are valid also for $L_T$) and $L_5$ which are sufficiently large to see the
hysteresis signal in the plaquettes and are therefore our estimates for the
minimal sizes $\Lsmin$ and $\L5min$. The dotted line simply represents
the couplings corresponding to $\gamma=1$.
\begin{figure}[t]
  \begin{center}
   \includegraphics*[angle=-90,width=.70\textwidth]{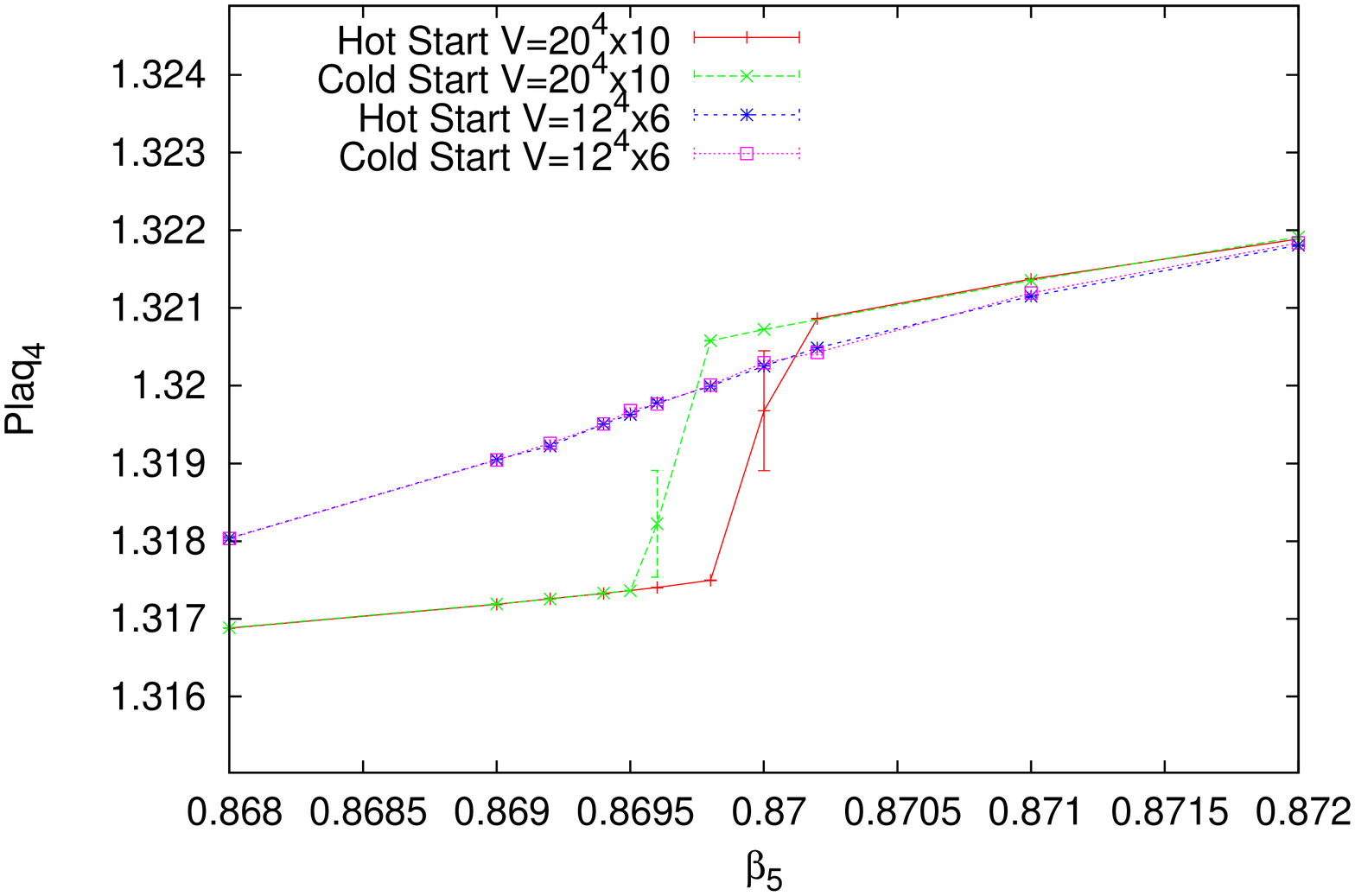}
\end{center}
  \caption{\small
Scan in $\beta_5$ of the plaquette $\plaq_4$ at $\beta_4=2.5$.
Lattices with $L_s$ and $L_T$ equal to 20 or larger are
needed to see the hysteresis.}
  \label{fig:asymm_bulk_fs}
\end{figure}

In \cite{Farakos:2010ie} it was claimed that at $\beta_4=3.0$ and
$\beta_5=0.779(1)$ the bulk phase
transition turns second order, as it was found in a mean-field computation 
\cite{Irges:2009bi,Irges:2009qp}. At $\beta_4=3.0$ lattices of much
larger size than the ones simulated so far are needed in order to determine
the order of the phase transition.
Given the large computational effort, at present we cannot comment on the
situation at $\beta_4=3.0$.
In \fig{fig:asymm_bulk_fs} we present our results at $\beta_4=2.5$.
We show the behavior of $\pq{4}$ as a function of $\beta_5$
for simulations with a hot start or a cold start. Each simulation consists of
$10^4$ measurements separated by 10 update iterations, composed by 1
heatbath and $\Ls/2$ overrelaxation sweeps each. We keep $L_T=\Ls$ and
$L_5=\Ls/2$. \fig{fig:asymm_bulk_fs} shows that for $\Ls=12$ (asterisks for
the hot and empty squares for the cold run) there is no hysteresis but a
smooth cross-over.
For $\Ls=20$ (plusses for the hot and crosses for the cold run) the
hysteresis appears indicating the presence of a bulk first order phase
transition. We list some values of the critical couplings for the bulk phase
transition at $\gamma<1$ in \tab{tab:1_asymbulk}. These couplings are chosen
to be approximately in the middle of the hysteresis region.
\tab{tab:1_asymbulk} also contains the lattice size $\Ls$ which is sufficiently
large to see the hysteresis and is our estimate of $\Lsmin$ in \eq{lmin}. If
$L_T$ or $\Ls$ is chosen smaller than $\Lsmin$ one sees a cross-over which is
due to ``compactification'' of the temporal or spatial dimension respectively. 

\subsection{Center breaking phase transitions \label{s_asymm_center}}

\begin{figure}[t]
  \begin{center}
   \includegraphics*[angle=-90,width=.80\textwidth]{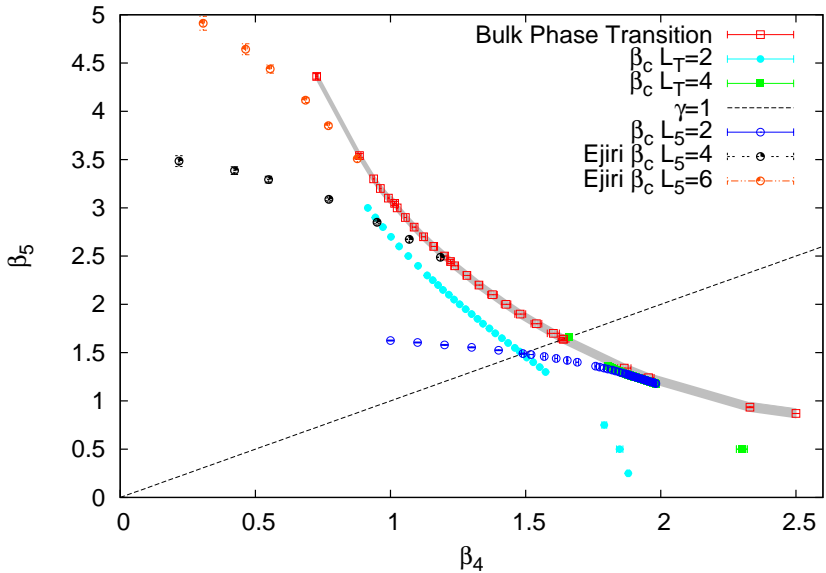}
\end{center}
  \caption{\small
The phase diagram of the five-dimensional SU(2) gauge theory with the
anisotropic Wilson plaquette action. Explanations are in the text.
}
  \label{fig:asymm_center_1}
\end{figure}
In \fig{fig:asymm_center_1} we complete our phase
diagram of five-dimensional SU(2) gauge theories in the $(\beta_4,\beta_5)$
coupling plane of the anisotropic Wilson plaquette gauge action.
In addition to the first order bulk phase transitions (empty squares),
studied in \sect{s_asymm_bulk}, we plot 
phase transitions due to center breaking (or
compactification) along either the temporal or the fifth dimension. The order
parameter is the Polyakov loop in the temporal or fifth dimension respectively.
Our new results are mainly at $\gamma<1$ (i.e. $\beta_4>\beta_5$, the region
below the dashed line in \fig{fig:asymm_center_1}).
For $\gamma>1$ (i.e. $\beta_4<\beta_5$, the region above the dashed line in
\fig{fig:asymm_center_1}) we include the points of \cite{Ejiri:2000fc}.
In \cite{Lang:1986kq,Ejiri:2000fc,deForcrand:2010be} the
transitions due to compactification of the fifth dimension have been studied
and found to be second order. These transitions are represented by empty
circles in \fig{fig:asymm_center_1}. For $L_5=2$ they
extend the $\gamma=1$ transition which we discussed in
\sect{sect:symmetric}. This line ends at $\gamma<1$ around $\beta_4=2$ when it
``hits'' the bulk phase transition line.
The transitions at $L_5=4$ and $L_5=6$ found in \cite{Ejiri:2000fc}
branch off the bulk phase transition line at $\gamma>1$ as $\beta_4$ is
lowered more and more.
\begin{table}[t]
\begin{center}
\begin{tabular}{c|c|c|c|c}
Volume & $\beta_{4c}$ ($\poly_T$) & $\beta_{4c}$ ($\poly_T(1)$) &
$\chi_c$ ($\poly_T$) & $\chi_c$ ($\poly_T(1)$)\\
\hline
$2\times8^3\times4$      & 1.83535(14)  & 1.83751(25) 
                         & 46.38(18)    & 16.69(7)    \\
$2\times12^3\times6$     & 1.82713(19)  & 1.82859(20) 
                         & 112.6(1.4)   & 29.3(3)     \\
$2\times16^3\times8$     & 1.82455(20)  & 1.82588(19) 
                         & 216(5)       & 44.4(9)     \\
$2\times20^3\times10$    & 1.82325(7)   & 1.82393(8)  
                         & 351(5)       & 57.4(6)     \\
$2\times24^3\times12$    & 1.82263(3)   & 1.82309(4)  
                         & 525(6)       & 70.9(5)     \\
\hline
$2\times\infty^4$        & 1.82115(8)   & 1.82095(19) & - & -
\end{tabular}
\end{center}
\caption{\small For the center
  breaking phase transition at $\beta_5=0.5$, $L_T=2$, we list
  the critical values of the coupling $\beta_{4c}$ and the
  susceptibility of the temporal Polyakov loop $\chi_c$ 
  as function of the volume. We use two definitions of the Polyakov line,
  \eq{polyT} and \eq{polyTx5}.}
\label{tab:1_asym_center}
\end{table}
\begin{figure}[t]
  \begin{center}
   \includegraphics*[angle=-90,width=.48\textwidth]{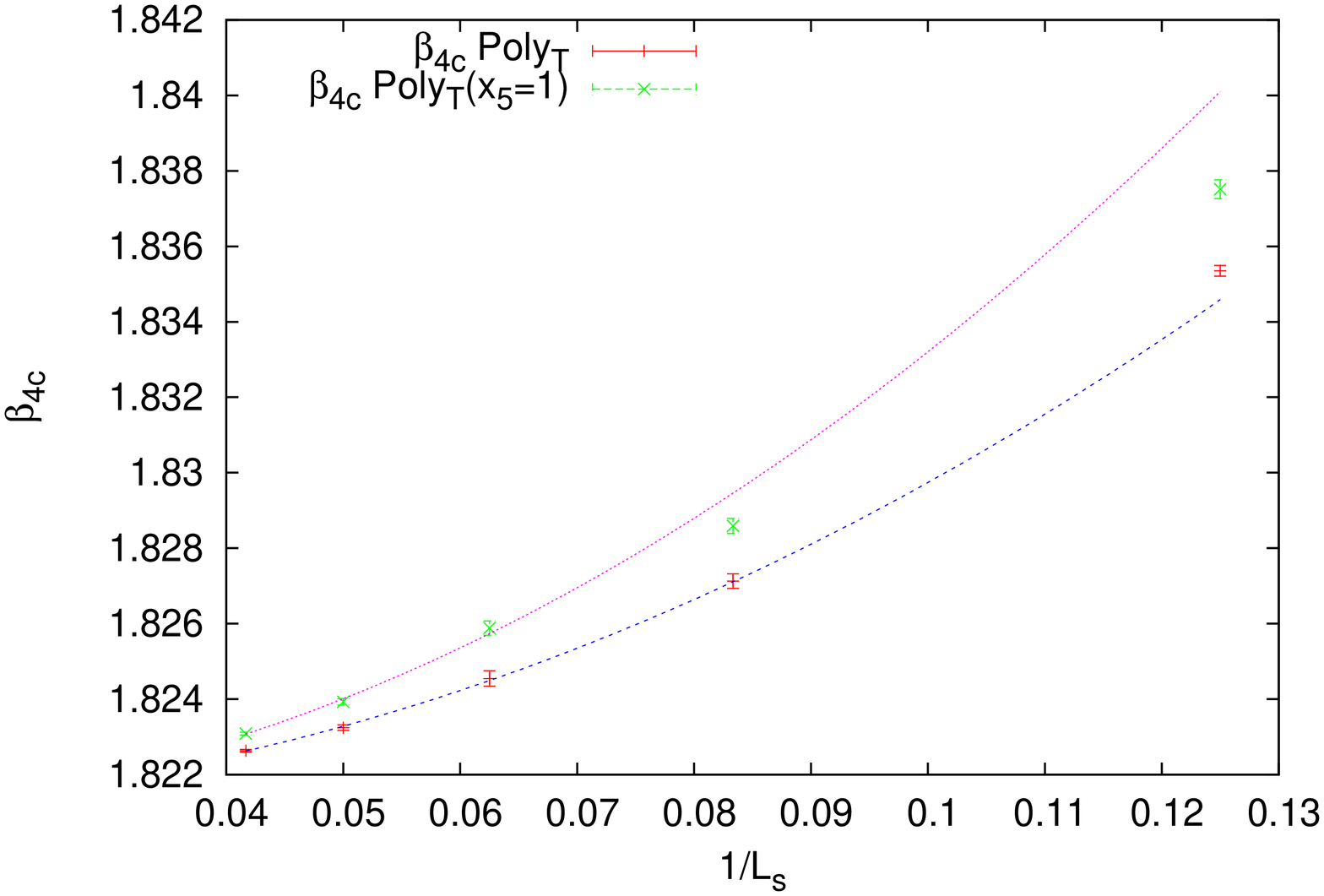}
   \includegraphics*[angle=-90,width=.48\textwidth]{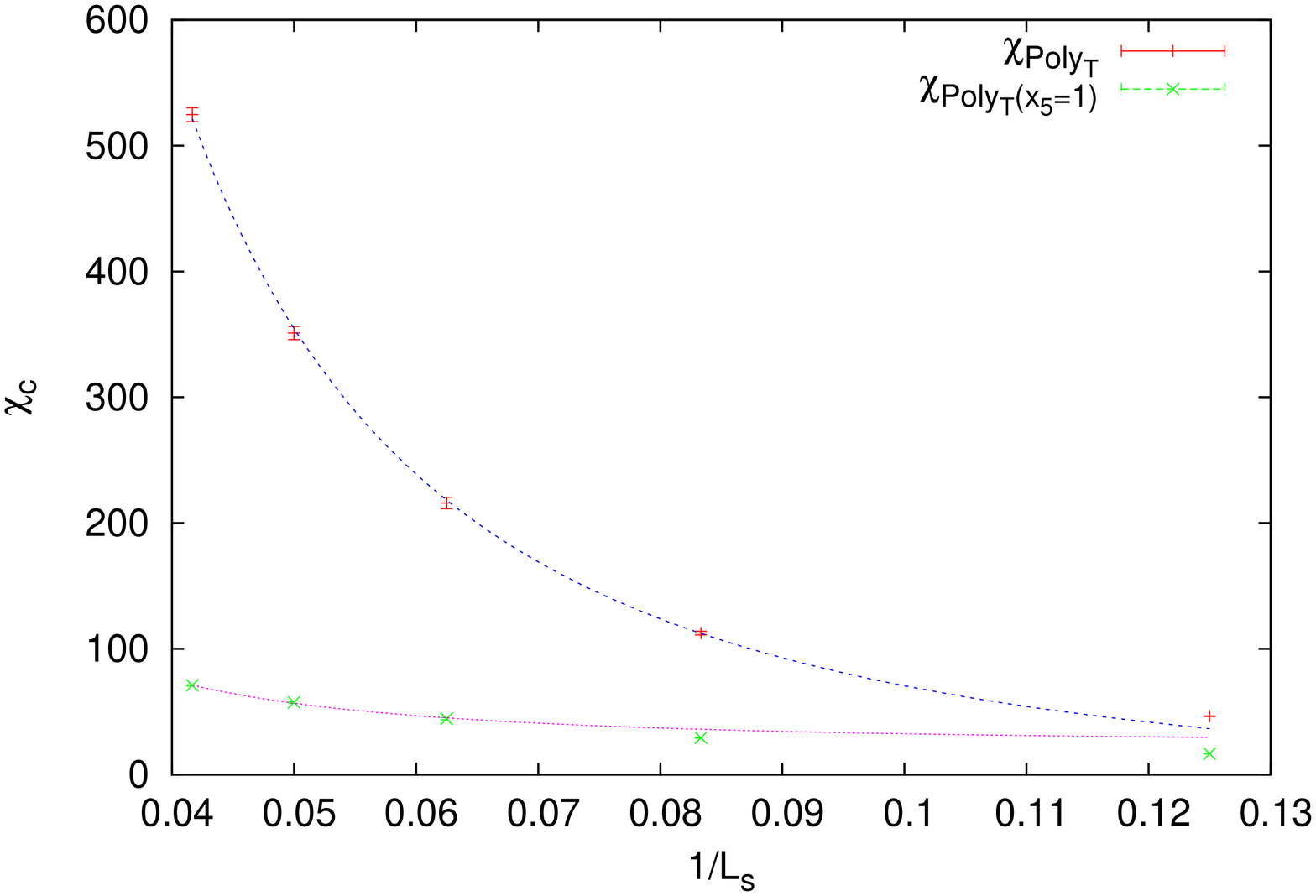}
\end{center}
  \caption{\small Finite size scaling analysis for the transition at
    $\beta_5=0.5$, $L_T=2$ based on \tab{tab:1_asym_center}.
    We use two definitions of the temporal Polyakov loop,
    \eq{polyT} (plusses) and \eq{polyTx5} (crosses).
    The lines are fits to the data using the critical exponents of the
    four-dimensional Ising model.}
  \label{fig:asymm_center_2}
\end{figure}
\begin{figure}[th!]
  \begin{center}
   \includegraphics*[angle=-90,width=.60\textwidth]{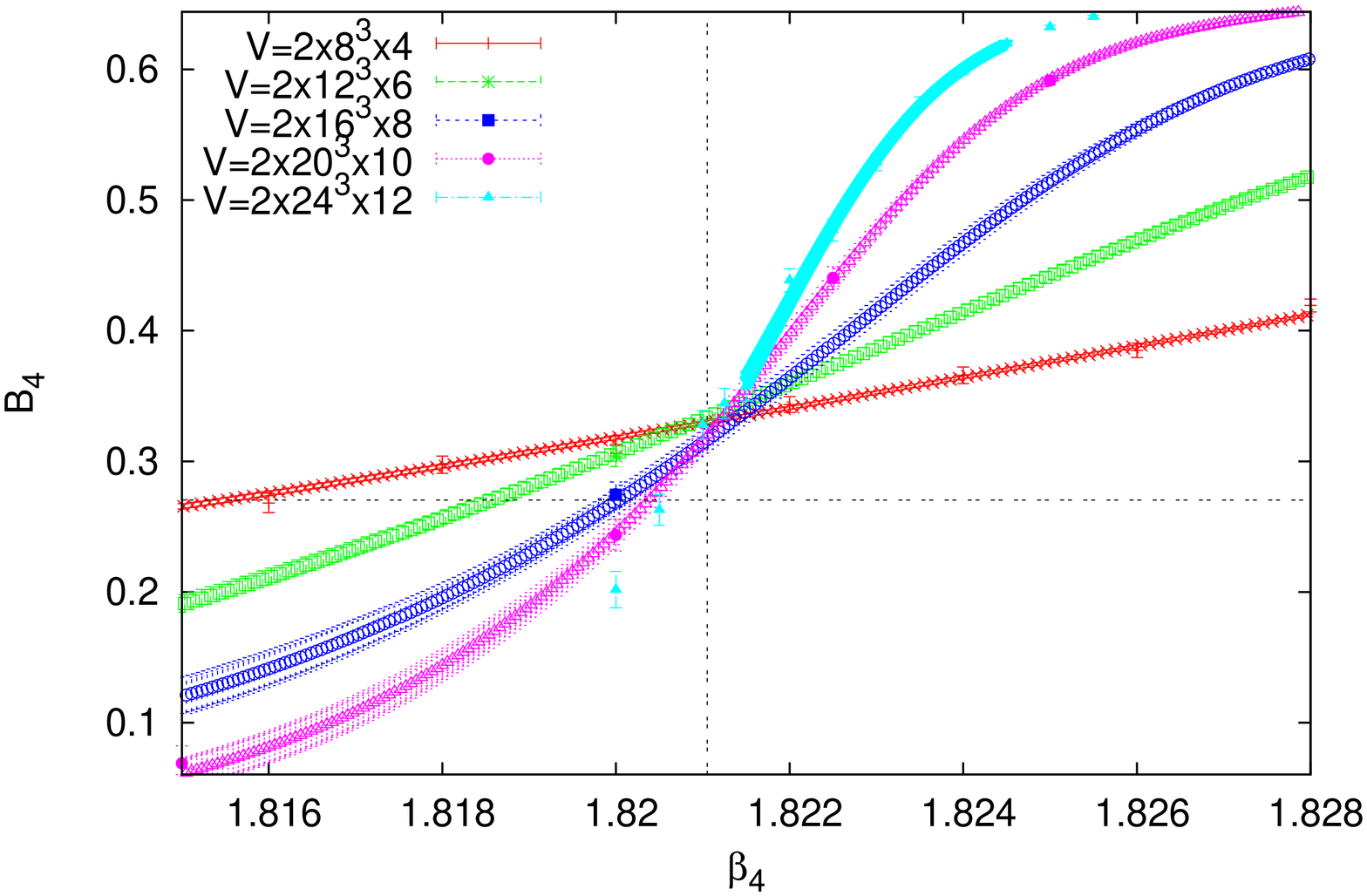}
\end{center}
  \caption{\small Finite size scaling analysis for the transition at
    $\beta_5=0.5$, $L_T=2$: the Binder
    cumulant $B_4$ of $\poly_T$ as a function of $\beta_4$ for several volumes.}
  \label{fig:asymm_center_3}
\end{figure}

In this article we study the transitions due to compactification of the
temporal dimension. For $L_T=2$ (filled circles in \fig{fig:asymm_center_1})
they are again continuation of the $\gamma=1$ transition. We performed a
finite size scaling analysis for the transition point at $\beta_5=0.5$.
In \tab{tab:1_asym_center} we list the critical values $\beta_{4c}$ of the
coupling $\beta_4$ at which the susceptibility of the temporal Polyakov line
has its maximum $\chi_c$ for several values of the lattice size
$\Ls=8\ldots24$ keeping $L_5=\Ls/2$. 
We use both definitions of the Polyakov line averaged along
the extra dimension \eq{polyT} or taken on a fixed slice $x_5=1$ \eq{polyTx5}.
In \fig{fig:asymm_center_2} we plot the data as function of $1/\Ls$ together
with fits using the critical exponents of the four-dimensional Ising model,
cf. \eq{2ndpt_1} and \eq{2ndpt_2} with $\beta_c$ replaced by $\beta_{4c}$ and
$L$ by $\Ls$. The fits
work quite well and using the largest three volumes we estimate the critical
value $\beta_{4c}(\Ls=\infty)=1.82115(8)$. Both definitions of the Polyakov
line give a perfectly compatible result.

In \fig{fig:asymm_center_3} we show the results for the Binder cumulant $B_4$
of the temporal Polyakov loop \eq{B4polyT} (here we use only the definition
of $\poly_T$ in \eq{polyT}). We apply the reweighting technique as discussed
in \sect{sect:symmetric} to get a denser data set. There is clear trend in the
data to intersect at a common value. The vertical line marks $\beta_{4c}$ and
the horizontal line the expected value for the universality class of the
four-dimensional Ising model \cite{deForcrand:2010be}. In order to precisely
determine the critical behavior larger volumes are needed. Please note that
the lattices which we simulated already require significant computational
resources.
\begin{table}
\begin{center}
\begin{tabular}{c|c|c}
Volume & $\beta_{4c}$ ($\poly_T(1)$) & $\chi_c$ ($\poly_T(1)$) \\
\hline
$4\times12^3\times6$     & 2.3024(8)   & 23.44(16) \\
$4\times16^3\times8$     & 2.2993(5)   & 41.1(3)   \\
$4\times20^3\times10$    & 2.2995(6)   & 62.2(5)   \\
$4\times24^3\times12$    & 2.2985(10)  & 88(3)     \\
$4\times32^3\times16$    & 2.2981(4)   & 156(2)    \\
\hline
$4\times\infty^4$        & 2.2974(4)   & -
\end{tabular}
\end{center}
\caption{\small For the center breaking phase transition at $\beta_5=0.5$,
  $L_T=4$, we list the
  critical values of the coupling $\beta_{4c}$ and the
  susceptibility of the temporal Polyakov loop $\chi_c$
  as function of the volume.
  We only use the definition of the Polyakov line \eq{polyTx5}.}
\label{tab:2_asym_center}
\end{table}
\begin{figure}[t]
  \begin{center}
   \includegraphics*[angle=0,width=.48\textwidth]{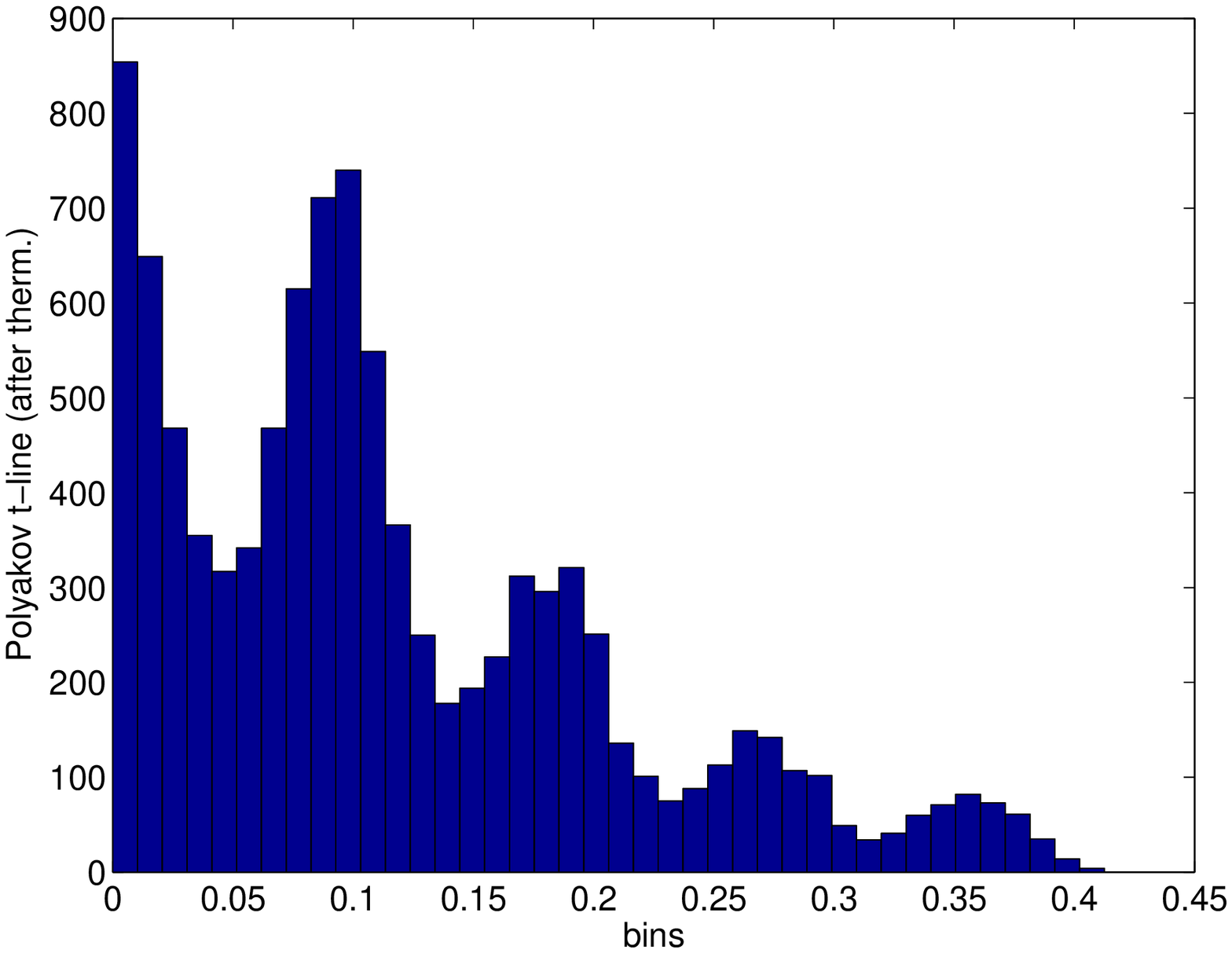}
   \includegraphics*[angle=0,width=.48\textwidth]{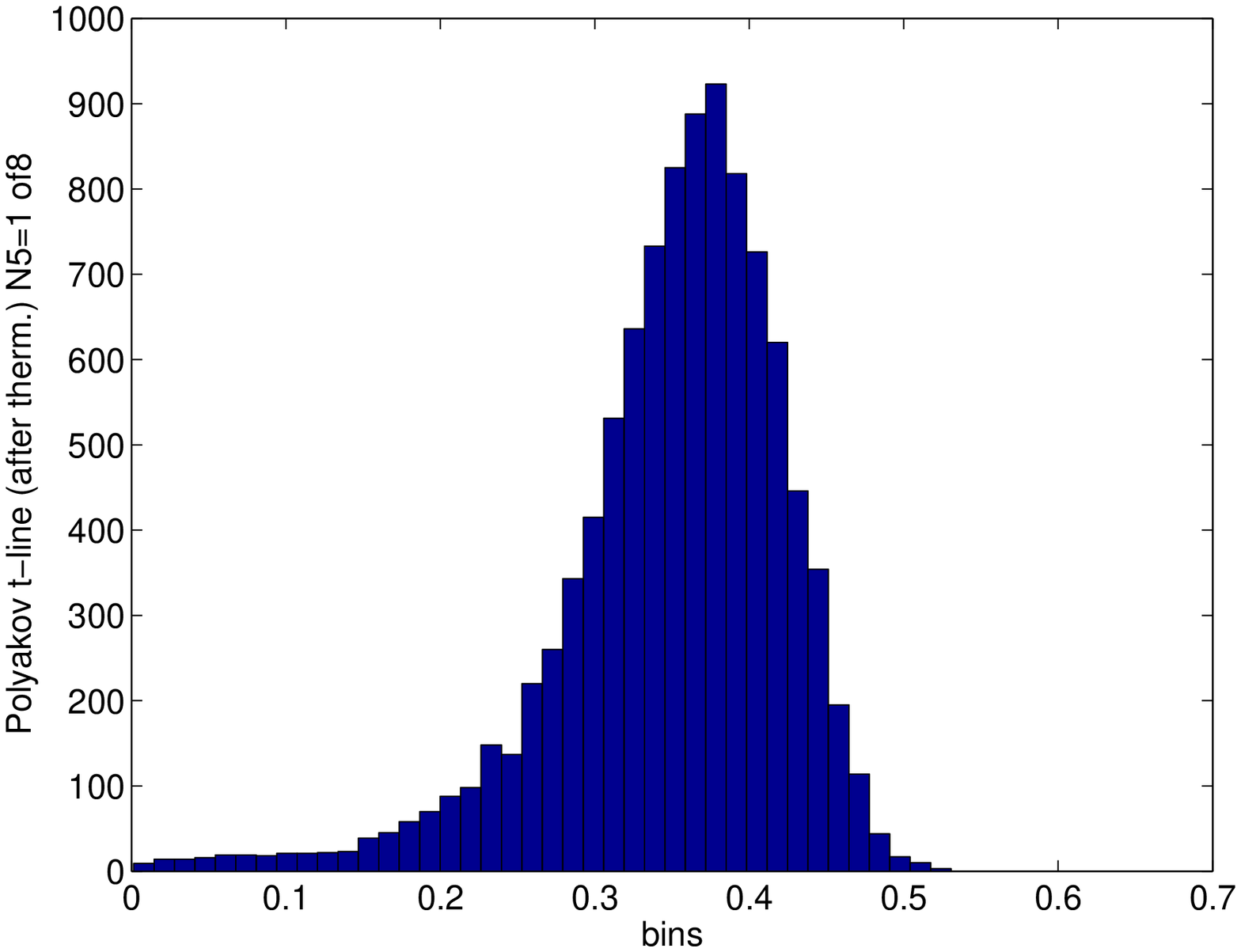}
\end{center}
  \caption{\small Histograms of the absolute value of the temporal Polyakov
    loop at $\beta_4=2.32$, $\beta_5=0.5$ and $L_T=4$: taking the definition
    \eq{polyT} (average over the extra dimension, left plot) or
    \eq{polyTx5} (at a fixed slice $x_5=1$, right plot).}
  \label{fig:asymm_center_4}
\end{figure}
\begin{figure}[th!]
  \begin{center}
   \includegraphics*[angle=-90,width=.48\textwidth]{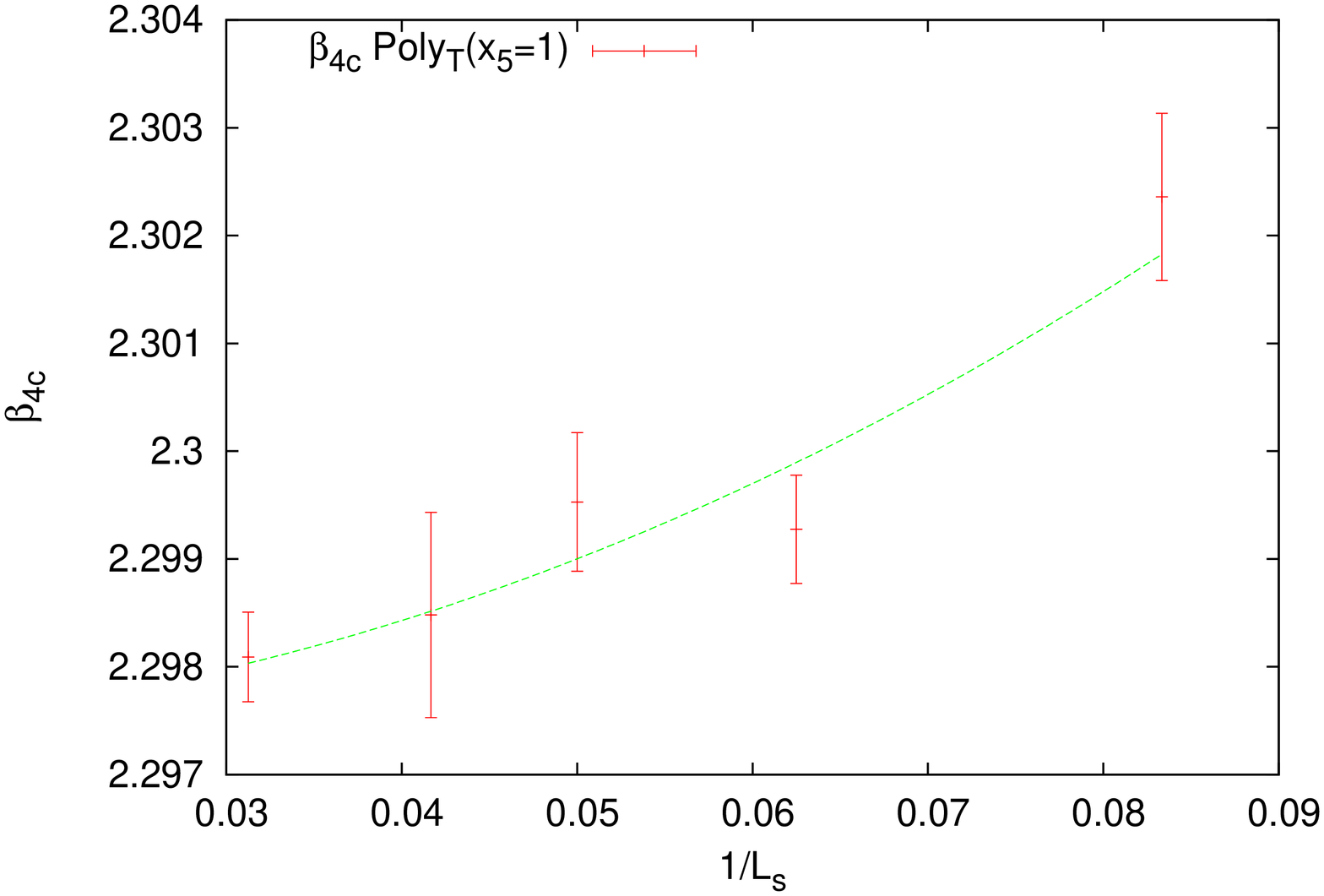}
   \includegraphics*[angle=-90,width=.48\textwidth]{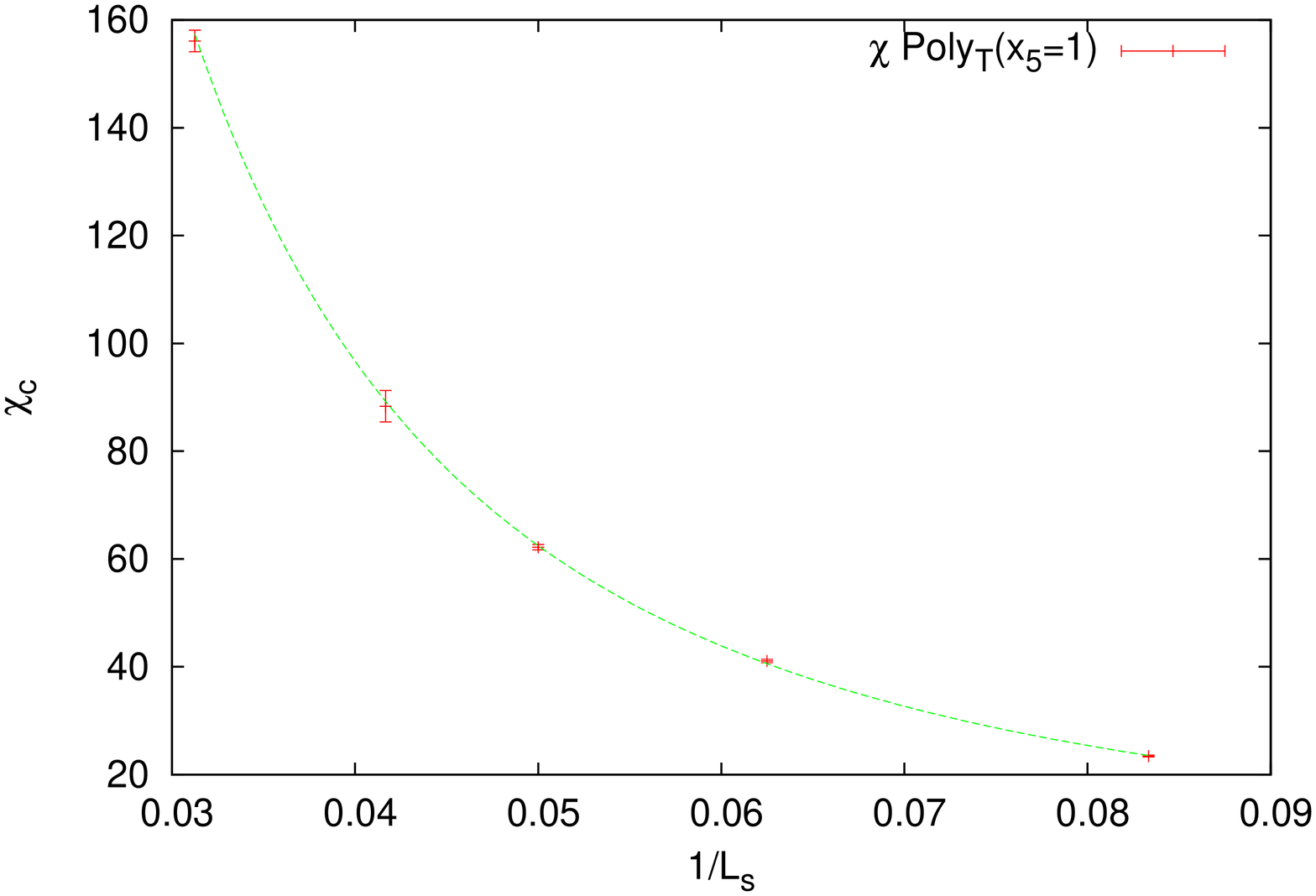}
\end{center}
  \caption{\small Finite size scaling analysis for the transition at
    $\beta_5=0.5$, $L_T=4$ based on \tab{tab:2_asym_center}.
    The temporal Polyakov line is taken in the slice $x_5=1$.
    The lines are fits to the data using the critical exponents of the
    four-dimensional Ising model.}
  \label{fig:asymm_center_5}
\end{figure}

For $L_T=4$ (filled squares in \fig{fig:asymm_center_1}) we faced a new
situation. Staying at $\beta_5=0.5$ (like for $L_T=2$), in the broken phase
close to the phase transition the histogram of the
temporal Polyakov loop \eq{polyT} shows several peaks.
This is shown by the plot on the left of
\fig{fig:asymm_center_4} for $\beta_4=2.32$.
If we do not average the Polyakov loop along the
fifth dimension and take its definition \eq{polyTx5} for the slice $x_5=1$,
we get the histogram shown by the plot on the right of
\fig{fig:asymm_center_4}. There is only a single peak. The interpretation of
these plots is that we are in a phase of the theory where the Polyakov lines
can fluctuate almost independently in the hyperplanes orthogonal to the fifth
dimension. In each hyperplane the distribution of their values has the
characteristic double-peak shape which appears like the plot on the right
of \fig{fig:asymm_center_4} when taking the absolute value. When averaging
over the hyperplanes they produce a multiple-peak structure, which is
therefore an artefact and does not mean the presence of multiple (more than
two) vacua. This is a very interesting effect signaling that the interactions
between the hyperplanes is weak. The decoupling of the hyperplanes for
$\gamma<1$ was in fact also found by the mean-field computation of
\cite{Irges:2009qp}. While the
geometrical setup of our computation is different, the qualitative feature we
see are the same.

Taking the temporal Polyakov line defined only on the slice $x_5=1$ we can do
a finite size scaling analysis as we previously did for $L_T=2$. 
It is now clear why we defined two Polyakov loop operators. Indeed in the
present case only for the one not averaged over the extra dimension a finite
size scaling analysis can be made. Where it was possible to perform both
analyses, we showed that their results coincided.
The critical values $\beta_{4c}$ and the maximum of the susceptibility $\chi_c$
are listed in \tab{tab:2_asym_center} and shown in
\fig{fig:asymm_center_5}. The fits to the critical behavior \eq{2ndpt_1} and
\eq{2ndpt_2} assuming the critical exponents of the four-dimensional Ising
model work well and give $\beta_{4c}(\Ls=\infty)=2.2974(4)$.

In summary the phase transitions due to compactification of the temporal
dimension at $\gamma<1$ for $L_T=2$ and $L_T=4$ are second order and
compatible with the critical exponents of the four-dimensional Ising model
thus confirming \cite{Svetitsky:1982gs}.

\subsection{The static potential \label{s_asymm_potential}}

\begin{figure}[t]
  \begin{center}
   \includegraphics*[angle=0,width=.70\textwidth]{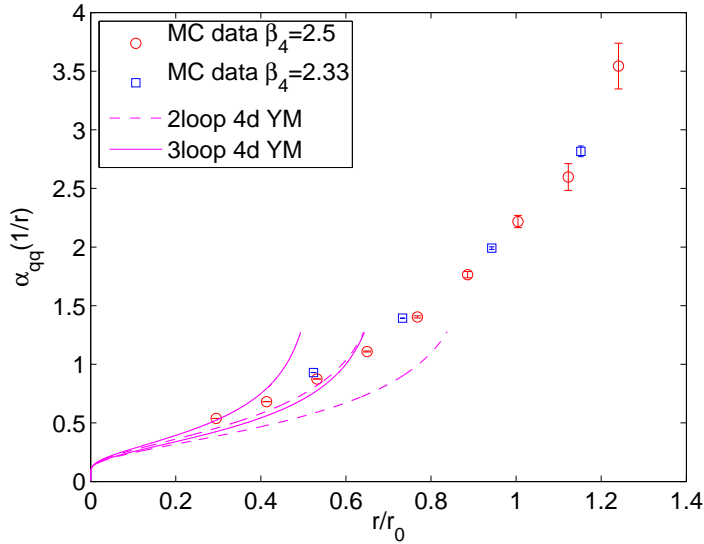}
\end{center}
  \caption{\small The coupling $\aqq$ \eq{aqq} at the bulk
    phase transition in the confined vacuum. The simulations have been
    performed at $\beta_4=2.33,\;2.5$, cf \tab{tab:1_asymbulk}.}
  \label{fig:asymm_pot1}
\end{figure}
The simulation results presented in the previous section show that at
$\gamma<1$ dimensional reduction from five to four dimensions can occur
by compactifying one of the directions orthogonal to the fifth dimension.
We discovered that the relevant degrees of freedom (the broken Polyakov loops)
fluctuate almost independently when defined in the hyperplanes orthogonal to
the extra dimension. The theory in the broken phase
reduces to four dimensions and not three, thereby indicating that the
hyperplanes do not exactly decouple but some interaction is left. Here we
go back to simulations of the theory in ``infinite volume''. By this we mean
that all five directions are larger than their minimal size, see \eq{lmin}. If
the hyperplanes decouple, the theory reduces to four dimensions, as
is the case in the mean-field computation of \cite{Irges:2009qp}.

We computed the static potential $V(r)$ along spatial directions orthogonal to
the extra dimension (and averaged over the extra dimension). In the
deconfined phase (where the Polyakov loops are broken) it is a
five-dimensional Coulomb potential \cite{Knechtli:2010sg}. Here we choose
parameters which correspond to bulk phase transition points and we put
ourselves approximately in the middle of the hysteresis curve.  We
started the simulations with a hot start in order to stay in the vacuum
of the confined phase and we checked that the Polyakov loop
expectation values are indeed zero in all directions. We choose the points in
parameter space $(\beta_4=2.33,\beta_5=0.937)$ and
$(\beta_4=2.5,\beta_5=0.8697)$, cf \fig{fig:asymm_bulk_fs},
and the lattice size is $32^4\times16$.
We ran 4 replica at $\beta_4=2.33$ and 11 replica at $\beta_4=2.5$ for a
total of 38280 and 103410 measurements respectively. Each
replicum used 512 cores of the supercomputer Cheops of the University of
Cologne and consists of $10^4$ update iterations (each iteration does one
heatbath and 16 overrelaxation sweeps) for thermalization and approximately
$10^4$ measurements of the Wilson loops separated by 10 update iterations.
We take four levels of HYP smearing for the spatial links of the Wilson loops.
We measure the potential starting from distance $r=2a_4$ and the force from
$r=2.5a_4$.
For the fit of the effective masses in \eq{fiteff} we use the range
$t=2,3,\ldots,6$.
The values of the scale $r_0/a_4$ are given in \tab{tab:1_asymbulk}. At
$\beta_4=2.5$ the lattice spacing $a_4$ is almost half the one at
$\beta_4=2.33$.
\begin{figure}[t]
  \begin{center}
   \includegraphics*[angle=0,width=.70\textwidth]{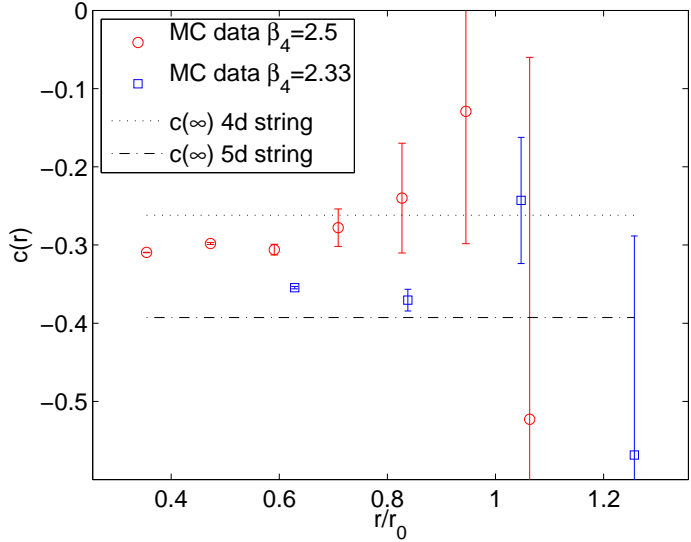}
\end{center}
  \caption{\small The slope $c$ \eq{ccoeff} at the bulk
    phase transition in the confined vacuum, from the same simulations as
    in \fig{fig:asymm_pot1}.}
  \label{fig:asymm_pot2}
\end{figure}

In \fig{fig:asymm_pot1} we show the coupling $\aqq$ \eq{aqq} as a function of
$r/r_0$, square symbols for $\beta_4=2.33$ and circles for $\beta_4=2.5$.
The string tension is estimated to be $\sigma r_0^2=1.3(2)$ and $\sigma
r_0^2=1.4(2)$ at $\beta_4=2.33$ and $\beta_4=2.5$ respectively.
We plot the 2-loop (dashed) and 3-loop (solid) perturbative curves for the
SU(2) Yang--Mills theory in four dimensions, as explained in Appendix
\ref{s_appa}. The points at the smallest distances are compatible with the
3-loop running, especially on the finer lattice at $\beta_4=2.5$.

\fig{fig:asymm_pot2} presents the analysis of the slope $c(r)$ \eq{ccoeff}.
At $\beta_4=2.33$ the values of $c$ seem to be more compatible with the
five-dimensional value of the L\"uscher coefficient (marked by a horizontal
dashed-dotted line).
As we lower the lattice spacing $a_4$, at $\beta_4=2.5$ the data have a clear
trend towards the four-dimensional value of the L\"uscher coefficient (marked
by a horizontal dotted line).
We loose the statistical signal above distance $r=r_0$. But as was noticed in
\cite{Luscher:2002qv} the onset of the string behavior in the static
potential starts already at distance $r_0$.
There is no scaling between the two simulations and we are most probably
not on a line of constant physics.
We interpret our results as an indication that increasing $\beta_4$ along the
bulk phase transition line makes the theory more four-dimensional than
five-dimensional, hinting at a dynamical localization mechanism for the gauge
field.

It is not clear whether there exists a continuum limit when $\beta_4$
increases while staying in the confined phase at $\gamma<1$. But even if it
does not exist, there could be a window of values of the lattice spacing for
which the cut-off effects are small and the five-dimensional gauge theory can
be described by an effective four-dimensional theory for energies much
smaller than the cut-off.

\section{Conclusions}
\label{sect:conclusions}

Lattice gauge theory is the tool to explore five-dimensional gauge theories
away from their trivial limit. So far we did not find in Monte Carlo
simulations of gauge group SU(2) using the anisotropic Wilson gauge action
and periodic boundary conditions
a second order bulk phase transition where a continuum limit could
be taken. We located a line of first order bulk phase transitions separating
the confined from the deconfined phase. In this article we studied in
particular the phase diagram for
anisotropy $\gamma<1$ (where the lattice spacing $a_5$ is larger than $a_4$).
As we move along the bulk phase transition line, while decreasing $a_4$ and
staying in the large volume confined phase,
we find indications of dimensional reduction.
Dimensional reduction at $\gamma<1$ was previously found in a mean-field
calculation \cite{Irges:2009qp} and relied on a decoupling of the hyperplanes
orthogonal to the extra dimension. This effect is seen in our analyses of
second order phase transitions related to breaking of the center. At these
transitions our finite size scaling studies are perfectly compatible with the
critical exponent of the four-dimensional Ising model.

As a next step we would like to measure the mass spectrum in the scalar and
vector channels in order to understand what exactly is the (almost)
four-dimensional theory that we get in the hyperplanes orthogonal to the extra
dimension at $\gamma<1$. These studies prepare the ground for future
simulations of the theory with orbifold boundary conditions
\cite{Irges:2004gy,Irges:2006hg,Ishiyama:2009bk}.

\bigskip

{\bf Acknowledgement.}
This work was funded by the Deutsche Forschungsgemeinschaft (DFG) under
contract KN 947/1-1. In particular A. R. acknowledges full support from the DFG.
We thank Nikos Irges and Bj\"orn Leder for their inputs during discussions.
The Monte Carlo simulations were carried out on the
Cheops supercomputer at the RRZK computing centre of the University of Cologne
and on the cluster Stromboli at the University of Wuppertal and we thank both
Universities.

\begin{appendix}
\section{Perturbation theory in the qq-scheme \label{s_appa}}

We summarize the steps needed to compute the 3-loop running of
\beq
\gqq^2(\mu) &=& \frac{4\pi}{\CF} r^2 F(r)\,,\;\mu=1/r \,,
\eeq
where $C_\mathrm{F}=(N^2-1)/(2N)$, in the SU(N) Yang--Mills theory
in four dimensions. The beta function is defined by
the renormalization group equation
\beq
  \mu\frac{{\rm d}}{{\rm d}\mu}\gqq(\mu) &=& \betaqq(\gqq(\mu)) \label{betafun}
\eeq
and has the perturbative expansion
\beq
\betaqq(\gqq) &=&
-\gqq^3 \left(\bqq_0 + \bqq_1 \gqq^2 + \bqq_2 \gqq^4 \ldots \right) \,.
\eeq
The coefficients ($\CA=N$)
\beq
\bqq_0 = b_0 & = & \frac{1}{(4\pi)^2}11\CA/3 \,, \label{b0} \\
\bqq_1 = b_1 & = & \frac{1}{(4\pi)^4}34\CA^2/3 \,, \label{b1}
\eeq
are universal. The 3-loop coefficient
\beq
\bqq_2 &=&
\frac{\CA^3}{(4\pi)^6}
\left(\frac{206}{3}+\frac{44\pi^2}{3}-\frac{11\pi^4}{12}+\frac{242}{9}\zeta(3)\right)
+ \left(\frac{\pi^2}{3}-4\right)b_0^3
\eeq
is determined combining results from \cite{pot:2loop1,pot:2loop2}.
In order to integrate \eq{betafun} we evaluate numerically
\beq
 \frac{\Lambda_\mathrm{qq}}{\mu}  &=&
  \left(b_0\gqq^2\right)^{-b_1/(2b_0^2)} {\rm e}^{-1/(2b_0\gqq^2)}
           \exp \left\{-\int_0^{\gqq} {\rm d}  x
          \left[\frac{1}{ \betaqq(x)}+\frac{1}{b_0x^3}-\frac{b_1}{b_0^2x}
          \right]
          \right\} \,,
\nonumber \\  \label{e_lambdapar}
\eeq
where for $\betaqq$ we insert the truncated perturbative expansion.

For the evaluation of \eq{e_lambdapar} we need to know the Lambda-parameter in
the qq-scheme. For gauge group SU(2) it can be estimated from the data of
the Schr\"odinger Functional (SF) coupling in Table 5 of \cite{Luscher:1992zx}. 
We integrate \eq{e_lambdapar} in the SF scheme using the $b_2^{\rm SF}$
coefficient determined in \cite{Luscher:1995nr,Narayanan:1995ex}.
We take the smallest coupling from \cite{Luscher:1992zx} and use
the determination of the $r_0$ scale in \cite{Sommer:1993ce}. After
converting to the qq-scheme \cite{Luscher:1992zx} we estimate
\beq
 \Lambda_\mathrm{qq}\,r_0 &=& 0.80(11)
\eeq
from the 3-loop running (which agrees with the 2-loop result).

\end{appendix}

\bibliography{phase_diag_5d}     
\bibliographystyle{h-elsevier}   

\end{document}